\documentclass[12pt]{elsarticle}
\usepackage[margin=2.5cm]{geometry}
\usepackage{lipsum}

\makeatletter
\def\ps@pprintTitle{%
   \let\@oddhead\@empty
   \let\@evenhead\@empty
   \def\@oddfoot{\reset@font\hfil\thepage\hfil}
   \let\@evenfoot\@oddfoot
}
\makeatother
\usepackage{multirow}
\usepackage{amsfonts,color,morefloats,pslatex}
\usepackage{amssymb,amsthm, amsmath,latexsym}
\usepackage{threeparttable}
\usepackage{amsfonts,color,morefloats,pslatex}
\usepackage{amssymb,amsthm, amsmath,latexsym}
\allowdisplaybreaks[4]
\biboptions{numbers,sort&compress}

\newtheorem{theorem}{Theorem}
\newtheorem{lemma}[theorem]{Lemma}
\newtheorem{proposition}[theorem]{Proposition}
\newtheorem{corollary}[theorem]{Corollary}

\newtheorem{definition}[theorem]{Definition}
\newtheorem{example}[theorem]{Example}
\newtheorem{question}[theorem]{Question}

\newtheorem{remark}[theorem]{Remark}

\newcommand{\one}{\mathbf 1}

\newcommand{\gf}{{\F}}

\newcommand{\Tr}{\operatorname{Tr}}
\newcommand{\wt}{\operatorname{wt}}
\newcommand{\Norm}{\operatorname{N}}

\newcommand{\F}{\mathbb{F}}

\newcommand{\bC}{{\mathbb{C}}}

\newcommand{\C}{\mathcal{C}}

\newcommand{\bc}{{\mathbf{c}}}
\newcommand{\bd}{{\mathbf{d}}}

\usepackage{lineno}
\usepackage{blindtext}
\usepackage{threeparttable}
 \usepackage{graphicx}
 \usepackage{arydshln}
\allowdisplaybreaks[4]

\begin{document}

\begin{frontmatter}



\title{Euclidean and Hermitian Self-Orthogonal Codes from Divisible Codes}

\tnotetext[fn1]{
This research was supported in part by the National Natural Science Foundation of China under Grant 12271059 and in part by the Natural Science Basic Research Program of Shaanxi (No. 2026JC-YXQN-030).
}

\author[]{Ziling Heng}
\ead{zilingheng@chd.edu.cn}
\author[]{Chenkun Xie\corref{cor}}
\ead{13095999956@163.com}

\cortext[cor]{Corresponding author}

\address{School of Science, Chang'an University, Xi'an 710064, China}




\begin{abstract}
Self-orthogonality and divisibility are two fundamental properties of
linear codes, and the relationship between them has long been of
interest. Huffman and Pless (2003) showed that every binary \(4\)-divisible
code is Euclidean self-orthogonal, that a ternary code is Euclidean
self-orthogonal if and only if it is \(3\)-divisible, and that a
quaternary code is Hermitian self-orthogonal if and only if it is
\(2\)-divisible. The general case remained open for a long time. In
2024, Li and Heng settled the odd-characteristic case, proving that
every \(p\)-divisible code over \(\F_{p^m}\), with \(p\) odd, that
contains the all-one vector is Euclidean self-orthogonal. In this paper
we treat the even-characteristic case: we prove that every
\(4\)-divisible code over \(\F_{2^m}\) containing the all-one vector is
Euclidean self-orthogonal. We also prove that, for every prime \(p\),
every \(p\)-divisible linear code over \(\F_{q^2}\), where \(q=p^m\),
that contains the all-one vector is Hermitian self-orthogonal. 
In addition, we generalize
the Hermitian criterion to the Galois self-orthogonal case, proving that any $p$-divisible linear code over $\gf_{q^r}$ that contains the all-one vector is $\ell$-Galois self-orthogonal for 
\(\ell\in\{1,\dots,r-1\}\). 
 These criteria make the
construction of self-orthogonal codes from divisible codes simple and
efficient. As applications, we recover the known Euclidean and
Hermitian self-orthogonality of generalized Reed--Muller codes and
obtain several infinite families of new Euclidean and Hermitian
self-orthogonal codes, which give rise to infinite families of quantum
codes with excellent parameters.
\end{abstract}


\end{frontmatter}


\section{Introduction}\label{sec1}
Let \(q\) be a prime power and let \(\F_q\) be the finite field with
\(q\) elements.  Denote by $\gf_q^*=\gf_q\setminus \{0\}$.An \([n,k,d]\) linear code \(\C\) over \(\F_q\) is a \(k\)-dimensional
linear subspace of \(\F_q^n\) whose minimum Hamming distance is
\[
d=\min\bigl\{\operatorname{wt}(\mathbf c):\ \mathbf 0\neq\mathbf c\in\C\bigr\},
\]
where \(\operatorname{wt}(\mathbf c)\) denotes the number of nonzero
coordinates of \(\mathbf c\). If there dose not exist an $[n,k,d+1]$ linear code over $\gf_q$, then 
an $[n,k,d]$ linear code over $\gf_q$ is said to be \emph{distance-optimal}. 
The \emph{weight distribution}  of \(\C\)
is the sequence \((A_0,A_1,\dots,A_n)\), where
\[
A_w=\#\bigl\{\mathbf c\in\C:\ \operatorname{wt}(\mathbf c)=w\bigr\}
\]
is the number of codewords of weight \(w\). In particular,
\(A_0=1\) and \(\sum_{w=0}^{n}A_w=q^{k}\).
The \emph{weight enumerator} of \(\C\) is defined by
\[
A_{\C}(x)=\sum_{w=0}^{n}A_w\,x^{w}.
\] 
Weight distributions of linear codes are an important research topic in coding theory, as they contain crucial information about the codes.

For
\(\mathbf u=(u_1,\dots,u_n)\) and
\(\mathbf v=(v_1,\dots,v_n)\in\F_q^n\), the Euclidean inner product is defined by
\[
\langle\mathbf u,\mathbf v\rangle_E=\sum_{i=1}^{n}u_iv_i\in\F_q,
\]
and the Euclidean dual of an $[n,k,d]$ linear code $\C$ over \(\F_q\)  is defined as
\[
\C^{\perp_E}
=\bigl\{\mathbf v\in\F_q^n\mid
\langle\mathbf u,\mathbf v\rangle_E=0
\ \text{for all }\mathbf u\in\C\bigr\}.
\]
If \(\C\subseteq\C^{\perp_E}\), then \(\C\) is called \emph{Euclidean
self-orthogonal}.  For
\(\mathbf x=(x_1,\dots,x_n),\mathbf y=(y_1,\dots,y_n)\in\F_{q^2}^n\),
the Hermitian inner product is defined by
\[
\langle\mathbf x,\mathbf y\rangle_H
=\sum_{i=1}^{n}x_i\overline{y_i},
\qquad\overline{y_i}=y_i^q,
\]
and the Hermitian dual of an $[n,k,d]$ linear code $\C$ over \(\F_{q^2}\) is defined as
\[
\C^{\perp_H}
=\bigl\{\mathbf y\in\F_{q^2}^n\mid
\langle\mathbf x,\mathbf y\rangle_H=0
\ \text{for all }\mathbf x\in\C\bigr\}.
\]
The code \(\C\) over \(\F_{q^2}\)  is called  \emph{Hermitian self-orthogonal} if
\(\C\subseteq\C^{\perp_H}\).

Self-orthogonal codes occupy a central place in coding theory because
they connect classical codes with several important structures.
Through the CSS construction and the Hermitian construction, both Euclidean and Hermitian self-orthogonal codes over finite fields yield quantum error-correcting codes, and this is the most common source of stabilizer codes \cite{Ketkar06,CRSS98}.  Euclidean self-orthogonal codes also have applications in the constructions
of lattices \cite{Wan} and LCD codes \cite{Massey}.  The supports of the minimum-weight codewords of some
 Euclidean self-orthogonal codes and of their  Euclidean duals form combinatorial
\(t\)-designs by the Assmus--Mattson theorem \cite{AM69, Ding, liheng, WangHeng24}.
It is therefore useful to have simple tests that can recognize self-orthogonality without computing a full generator matrix.

A linear code \(\C\) is said to be \(\Delta\)-\emph{divisible} if
\(\Delta\mid\wt(\mathbf c)\) for every \(\mathbf c\in\C\). If \(\gcd(\Delta,q) = 1\), one can construct a code equivalent to \(\mathcal{C}\) by expanding a linear code over \(\mathbb{F}_q\) (repeating every coordinate \(\Delta\) times) and then appending zeros until its length matches that of \(\mathcal{C}\) \cite{Ward1}. Consequently, the most noteworthy scenario occurs when \(\Delta\) equals a power of the characteristic of \(\mathbb{F}_q\). Divisible codes were first introduced by Ward in 1981 \cite{Ward1}, followed by a survey in 2001 \cite{Ward2}. Subsequently, in 2020, Kiermaier and Kurz characterized the effective lengths for all \(q^r\)-divisible linear codes over \(\mathbb{F}_q\), where \(r\) is a non-negative integer \cite{KK}.

Divisibility is one of the structural properties most closely tied to
self-orthogonality. In 2003, Huffman and Pless showed that every binary
\(4\)-divisible linear code is Euclidean self-orthogonal, that a ternary
linear code is Euclidean self-orthogonal if and only if it is
\(3\)-divisible, and that a quaternary linear code is Hermitian
self-orthogonal if and only if it is \(2\)-divisible \cite{HP}.
For general \(q\)-ary codes, however, no analogous weight-only criterion
is known, and verifying self-orthogonality directly is usually difficult
\cite{WangHeng24}. A recent systematic study of self-orthogonal codes
arising from \(p\)-divisible codes appears in \cite{liheng}. For
\(q=p^m\) with \(p\) an odd prime, Li and Heng proved that every
\(q\)-ary linear code that contains the all-one vector and is
\(p\)-divisible is Euclidean self-orthogonal. The following questions
therefore remain open.

\begin{question}
When \(q=2^m\), what is the relationship between the Euclidean
self-orthogonality and the divisibility of a \(q\)-ary linear code?
\end{question}

\begin{question}
When \(q=p^m\) for a prime \(p\), what is the relationship between the
Hermitian self-orthogonality and the divisibility of a linear code over
\(\mathbb F_{q^2}\)?
\end{question}

In this paper, we answer both questions under the additional hypothesis
that the code contains the all-one vector. The answers are two clean
divisibility criteria for self-orthogonality: a Euclidean criterion in
even characteristic (Section~III), and a Hermitian criterion valid in
every characteristic (Section~IV). These are the main results of the
paper and are stated next.

\noindent\textbf{Euclidean criterion (Theorem~\ref{thm:code}).}
Let \(q=2^m\) with \(m\) a positive integer, and let
\(\C\subseteq\F_q^n\) be a linear code containing the all-one vector
\(\one\). If \(\C\) is \(4\)-divisible, then \(\C\subseteq\C^{\perp_E}\).

\medskip
\noindent\textbf{Hermitian criterion (Theorem~\ref{thm:main}).}
Let \(q=p^m\), where \(p\) is prime and \(m\) is a positive integer, and
let \(\C\subseteq\F_{q^2}^n\) be a linear code containing \(\one\). If
\(\C\) is \(p\)-divisible, then \(\C\subseteq\C^{\perp_H}\).

The Euclidean criterion is proved by a double-counting argument in the
affine plane over \(\F_q\), based on congruence conditions on the sums
of symbol counts along affine lines; the Hermitian criterion follows
from a symbol-count congruence combined with a polarization over
\(\F_{q^2}\).  In addition, we generalize
the Hermitian criterion to the Galois self-orthogonal case, proving that any $p$-divisible linear code over $\gf_{q^r}$ that contains the all-one vector is $\ell$-Galois self-orthogonal for 
\(\ell\in\{1,\dots,r-1\}\).

Both criteria admit \textbf{monomial-equivalence versions}: 
\begin{enumerate}
\item[$\bullet$] Every
\(4\)-divisible linear code of length $n$ over \(\F_q\) \((q=2^m)\) containing a codeword of
weight \(n\) is monomially equivalent to a Euclidean self-orthogonal
code (Corollary~\ref{cor:euclidean-monomial}).
\item[$\bullet$] Every \(p\)-divisible
linear code of length $n$ over \(\F_{q^2}\) \((q=p^m)\) containing a codeword of weight \(n\)
is monomially equivalent to a Hermitian self-orthogonal code
(Corollary~\ref{cor:monomial}). 
\end{enumerate}
The hypothesis \(\one\in\C\) in both theorems  cannot be
dropped when \(q>2\), as Remarks~\ref{rem:allone} and
\ref{rem-allone} show. The two criteria also recover
the known Euclidean and Hermitian self-orthogonality of generalized
Reed--Muller codes; see Remarks~\ref{rem:grm} and
\ref{rem:grm-hermitian}. Furthermore, the two criteria yield several infinite families of new Euclidean and Hermitian self-orthogonal codes. The Euclidean and Hermitian self-orthogonal codes constructed herein give rise to several infinite families of quantum codes with excellent parameters.

The remainder of the paper is organized as follows.  Section 2
collects the notation and the preliminary results used throughout.
Section 3 proves the Euclidean criterion and its monomial-equivalence corollary.
Section 4 proves the Hermitian criterion and its monomial-equivalence corollary.  Section
5 constructs the new families of self-orthogonal codes and studies
their parameters, weight enumerators, designs and locality.  Section
VI derives the quantum codes and discusses their optimality.

\section{Preliminaries}
In this section, we recall some basic definitions and facts that will be used in the
sequel.

\subsection{Trace and norm functions over finite Fields and properties of the Hermitian inner product}

\begin{definition}[Trace and norm functions]\label{def:trace-norm}
Let \(q\) be a prime power and \(r\ge1\). For \(x\in\F_{q^r}\),
the trace function from $\gf_{q^2}$ to $\gf_q$  is defined by
\[
\Tr_{q^r/q}(x)=\sum_{i=0}^{r-1}x^{q^i}\in\F_q.
\]
 For \(x\in\F_{q^r}^*\), the norm function from \(\F_{q^r}^*\) to \(\F_q^*\) is defined by
$$ \Norm_{q^r/q}(x)=\prod_{i=0}^{r-1}x^{q^i}=x^{(q^r-1)/(q-1)}\in\F_q^*.$$
\end{definition}

In particular, \(\Norm_{q^2/q}(x)=x^{q+1}\).
The trace function is \(\F_q\)-linear, whereas the norm function is multiplicative.
For \(t\mid r\),
\[
\Tr_{q^r/q}
=\Tr_{q^t/q}\circ\Tr_{q^r/q^t},
\qquad
\Norm_{q^r/q}
=\Norm_{q^t/q}\circ\Norm_{q^r/q^t}.
\]
In particular, \(\Tr_{q/p}\) with \(q=p^m\) is called the absolute trace function
from \(\F_q\) to \(\F_p\).

The following properties of the Hermitian inner product are well known
and can be easily verified from the definition.

\begin{lemma}\label{lem:hermitian}
For \(\mathbf{x},\mathbf{y},\mathbf{z}\in\F_{q^2}^n\) and
\(a,b\in\F_{q^2}\), the following properties hold.
\begin{enumerate}
\item[(i)]
\(
\langle \mathbf{x}+\mathbf{y},\mathbf{z}\rangle_H
=\langle \mathbf{x},\mathbf{z}\rangle_H
+\langle \mathbf{y},\mathbf{z}\rangle_H,
\langle \mathbf{x},\mathbf{y}+\mathbf{z}\rangle_H
=\langle \mathbf{x},\mathbf{y}\rangle_H
+\langle \mathbf{x},\mathbf{z}\rangle_H.
\)

\item[(ii)]
\(
\langle a\mathbf{x},\mathbf{y}\rangle_H
=a\langle \mathbf{x},\mathbf{y}\rangle_H,
\langle \mathbf{x},a\mathbf{y}\rangle_H
=\overline{a}\langle \mathbf{x},\mathbf{y}\rangle_H.
\)
In particular,
\(
\langle a\mathbf{x},b\mathbf{y}\rangle_H
=a\overline{b}\,\langle \mathbf{x},\mathbf{y}\rangle_H.
\)

\item[(iii)]
\(
\langle \mathbf{x},\mathbf{y}\rangle_H
=\overline{\langle \mathbf{y},\mathbf{x}\rangle_H}.
\)

\item[(iv)]
\(
\langle \mathbf{x},\mathbf{x}\rangle_H
=\sum_{i=1}^{n}x_{i}^{q+1}
=\sum_{i=1}^{n}\Norm_{q^{2}/q}(x_{i})\in\F_q,
\)
where \(\Norm_{q^{2}/q}\) is the norm from \(\F_{q^2}\) to
\(\F_q\), as defined in Definition~\ref{def:trace-norm}.

\item[(v)]
If \(\langle \mathbf{x},\mathbf{y}\rangle_H=0\) for all
\(\mathbf{y}\in\F_{q^2}^n\), then
\(\mathbf{x}=\mathbf{0}\).

\item[(vi)]
\(\langle \mathbf{x}+\mathbf{y},\mathbf{x}+\mathbf{y}\rangle_H
=\langle \mathbf{x},\mathbf{x}\rangle_H
+\langle \mathbf{y},\mathbf{y}\rangle_H
+\langle \mathbf{x},\mathbf{y}\rangle_H
+\langle \mathbf{y},\mathbf{x}\rangle_H.
\)
\end{enumerate}
\end{lemma}

\subsection{Finite geometry: lines and caps}

In this subsection, we collect some standard finite-geometric notions
and one classical bound that will be used in the sequel. We first recall
affine lines in the plane \(\F_q^2\), which will be used to express the
divisibility condition as congruences on line sums; we then recall caps
in projective spaces and Barlotti's cap bound.

\begin{definition}[Affine line]
An affine line \(L\subseteq\F_q^2\) is a set of the form
\[
L=\{(x,y)\in\F_q^2\mid ax+by=c\},
\]
where \(a,b,c\in\F_q\) and \((a,b)\neq(0,0)\).
\end{definition}

Every affine line of \(\F_q^2\) has exactly \(q\) points.

\begin{definition}[Projective space and cap]
The projective space \(\mathrm{PG}(r,q)\) consists of the
one-dimensional \(\F_q\)-subspaces of \(\F_q^{r+1}\). Three points are
collinear if their representatives lie in a common two-dimensional
subspace. A cap is a subset of \(\mathrm{PG}(r,q)\) containing no
three collinear points.
\end{definition}

\begin{lemma}[{Barlotti cap bound \cite[Theorem~9]{Barlotti}}]
\label{lem:barlotti-cap}
Let \(q>2\) be even and \(r\ge5\). If
\(\mathcal K\subseteq\mathrm{PG}(r,q)\) contains no three collinear points,
then \(|\mathcal K|<q^{r-1}\).
\end{lemma}

\subsection{Coding theory: codes, designs and locality}

In this subsection, we recall the coding-theoretic notions and bounds
that will be used in the sequel: monomial equivalence, \(t\)-designs
and the Assmus--Mattson theorem, and locally recoverable codes together
with the Singleton-like bound. Throughout, \(\C^{\perp_E}\) denotes the
Euclidean dual of \(\C\), and \(d(\C)\) its minimum distance.

\begin{definition}[Monomial equivalence]
Two linear codes \(\C,\C'\subseteq\F_q^{n}\) are \emph{monomially
equivalent} if there exist a permutation \(\pi\) of \(\{1,\dots,n\}\)
and scalars \(\lambda_{1},\dots,\lambda_{n}\in\F_q^{*}\) such that
\[
\C'
=\Bigl\{
\bigl(\lambda_{1}c_{\pi(1)},\dots,\lambda_{n}c_{\pi(n)}\bigr)\mid
\mathbf{c}=(c_{1},\dots,c_{n})\in\C
\Bigr\}.
\]
Equivalently, \(\C'=\C M\) for a monomial matrix
\(M=PD\), where \(P\) is a permutation matrix and
\(D=\operatorname{diag}(\lambda_{1},\dots,\lambda_{n})\) is an
invertible diagonal matrix.
\end{definition}

\begin{definition}[\(t\)-designs]\label{def:t-design}
Let \(v,k,t,\lambda\) be integers with \(0\le t\le k\le v\).
A \emph{\(t\)-\((v,k,\lambda)\) design} is a pair
\((X,\mathcal B)\), where \(X\) is a \(v\)-element set and
\(\mathcal B\) is a collection of \(k\)-element subsets of \(X\),
called blocks, such that every \(t\)-element subset of \(X\) is
contained in exactly \(\lambda\) blocks.
\end{definition}

A family of codewords of length \(v\) in a linear code $\C$ is said to \emph{support a
\(t\)-design} if every \(t\)-subset of the coordinate set is contained
in the same number of supports of these codewords, counted with
multiplicity; the corresponding support sets are then the blocks of a
\(t\)-design.

\begin{lemma}[{Assmus--Mattson Theorem \cite{AM69}}]
\label{lem:assmus-mattson}
Let \(\C\subseteq\F_q^n\) be a linear code with minimum distance \(d\),
and let \(d^{\perp}\) be the minimum distance of \(\C^{\perp_E}\). Let
\(t,t'\) be integers with \(0\le t\le d\) and \(0\le t'\le d^{\perp}\).

\begin{enumerate}
\item[(i)] If \(\C^{\perp_E}\) has at most \(d-t\) distinct nonzero
weights in \(\{1,2,\ldots,n-t\}\), then for every weight \(w\) with
\(d\le w\le n-t\) and \(A_w(\C)\neq0\), the supports of the codewords
of \(\C\) of weight \(w\) form a \(t\)-design.

\item[(ii)] If \(\C\) has at most \(d^{\perp}-t'\) distinct nonzero
weights in \(\{1,2,\ldots,n-t'\}\), then for every weight \(w\) with
\(d^{\perp}\le w\le n-t'\) and \(A_w(\C^{\perp_E})\neq0\), the supports
of the codewords of \(\C^{\perp_E}\) of weight \(w\) form a
\(t'\)-design.
\end{enumerate}

Here \(A_w(\C)\) denotes the number of codewords of weight \(w\) in a
code \(\C\).
\end{lemma}

\begin{definition}[Locality]
Let \(\C\subseteq\F_q^n\) be an $[n,k,d]$ linear code with a generator matrix
\(G=[\mathbf{g}_1,\dots,\mathbf{g}_n]\). The locality of the
\(i\)-th symbol is the smallest nonnegative integer \(\ell\) such that
\(\mathbf{g}_i\) is a linear combination of at most \(\ell\) of the
other columns of \(G\). If the locality is at most \(\ell\) for every
symbol, then \(\C\) is called an \((n,k,d,q;\ell)\)-locally recoverable code (LRC) with
locality \(\ell\).
\end{definition}

\begin{lemma}\cite[Cor.~3]{Tan23}
\label{lem:dual-design-locality}
Let \(\C\subseteq\F_q^n\) be a nontrivial linear code and let
\(d^{\perp}=d(\C^{\perp_E})\). If the supports of the minimum-weight
words of \(\C^{\perp_E}\) form a \(1\)-\((n,d^{\perp},\lambda)\) design
with \(\lambda\ge1\), then \(\C\) has locality \(d^{\perp}-1\).
\end{lemma}

\begin{lemma}[{Singleton-like bound \cite{Gopalan12}}]
\label{lem:lrc-singleton}
For an \((n,k,d,q;\ell)\)-LRC,
\[
d\le n-k-\left\lceil\frac{k}{\ell}\right\rceil+2.
\]
\end{lemma}

Following \cite{liheng}, an LRC is called \emph{distance-optimal} if
equality holds in the Singleton-like bound, and \emph{almost
distance-optimal} if its distance is one less than the right-hand side
of that bound.

\section{Euclidean self-orthogonal codes from divisible codes}
Throughout this section, let $q=2^m$ and $\C\subseteq\F_q^n$ be a linear code over $\F_q$ such that $\one=(1,\ldots,1)\in \C$ and  $\C$ is $4$-divisible. 
We prove $\C\subseteq \C^{\perp_E}$ in this section. To this end, we need the following lemmas.

For $\alpha\in\F_q^*$, define the $\F_2$-linear
function
\begin{equation*}
\ell_\alpha(z)=\Tr_{q/2}(\alpha z)\text{ for }z\in \F_q,
\end{equation*}
where $\Tr_{q/2}$ is the absolute trace from $\F_q$ to $\F_2$. 
Define
\begin{equation*}
W=\{u^2+u\mid u\in\F_q\}\subseteq\F_q
\end{equation*}
and
\begin{equation*}
W^\perp
=\left\{z\in\F_q\mid
\Tr_{q/2}(zw)=0\ \text{for all }w\in W
\right\}.
\end{equation*}

\begin{lemma}\label{lem:W-properties}
Let \(q=2^m\ge4\). Then the following assertions hold.
\begin{enumerate}
\item \( W=\ker(\Tr_{q/2})\).
\item every \(w\in W\) has exactly two preimages under the map
\(u\mapsto u^2+u\).
\item \(W^\perp=\F_2\).
\end{enumerate}
\end{lemma}

\begin{proof}
Since the absolute trace is \(\F_2\)-linear and invariant
under the Frobenius map,
\begin{equation*}
\Tr_{q/2}(u^2+u)
=\Tr_{q/2}(u^2)+\Tr_{q/2}(u)
=\Tr_{q/2}(u)+\Tr_{q/2}(u)=0.
\end{equation*}
Hence \(W\subseteq(\ker\Tr_{q/2})\). The polynomial
\(u^2+u=u(u+1)\) has exactly the two roots \(0\) and \(1\).
Thus the map \(u\mapsto u^2+u\) has kernel of size \(2\) and image
of size \(q/2\). On the other hand, the trace map is a nonzero
\(\F_2\)-linear map, so its kernel also has size \(q/2\).
Consequently,
\begin{equation*}
W=\ker(\Tr_{q/2}).
\end{equation*}

If \(u^2+u=w\), then
\begin{equation*}
(u+1)^2+(u+1)
=u^2+1+u+1
=u^2+u
=w.
\end{equation*}
Conversely, if \(u_1^2+u_1=u_2^2+u_2\), then
\begin{equation*}
(u_1+u_2)^2+(u_1+u_2)=0.
\end{equation*}
Hence \(u_1+u_2\in\{0,1\}\), and therefore
\(u_2\in\{u_1,u_1+1\}\). This proves the second assertion.

Finally, \(W=\ker(\Tr_{q/2})\) has dimension \(m-1\). Since the
trace pairing is nondegenerate, \(W^\perp\) has dimension \(1\).
Moreover, every \(w\in W=\ker(\Tr_{q/2})\) satisfies
\begin{equation*}
\Tr_{q/2}(1\cdot w)=0,
\end{equation*}
so \(1\in W^\perp\). Therefore
\begin{equation*}
W^\perp=\F_2.
\end{equation*}
This completes the proof.
\end{proof}

\begin{lemma}\label{lem:zero-trace-lines}
Let \(q=2^m\ge4\) and \(\alpha\in\F_q^*\). For \(a,b\in\F_q\),
define the nonvertical affine line
\begin{equation*}
L_{a,b}
=\{(s,t)\in\F_q^2\mid t=as+b\}.
\end{equation*}
Then \(\ell_\alpha(st)\) vanishes identically on \(L_{a,b}\) if and
only if
\begin{equation}\label{eq:zero-lines}
(a,b)=(0,0)
\quad\text{or}\quad
\bigl(b\neq0,\ a=\alpha b^2\bigr).
\end{equation}
\end{lemma}

\begin{proof}
If \(a=0\), then \(st=bs\). For \(b\neq0\), multiplication by
\(b\) is a permutation of \(\F_q\), and hence
\(\ell_\alpha(bs)\) is not identically zero as \(s\) varies.
Thus the required vanishing occurs exactly when \(b=0\), that is,
on the horizontal line \(t=0\).

If \(a\neq0\) and \(b=0\), then \(st=as^2\). Since
\(s\mapsto s^2\) is a permutation of \(\F_q\) and
\(a\neq0\), the values \(as^2\) run through all of \(\F_q\).
Therefore \(\ell_\alpha(st)\) is not identically zero on this line.

Now suppose that \(a,b\in\F_q ^*\). Put
\(s=(b/a)u\). Then
\begin{equation*}
st=s(as+b)=as^2+bs=\frac{b^2}{a}(u^2+u).
\end{equation*}
By Lemma~\ref{lem:W-properties},
\(\ell_\alpha(st)\) vanishes identically on \(L_{a,b}\) if and only if
\begin{equation*}
\ell_\alpha\!\left(\frac{b^2}{a}w\right)=0
\quad\text{for every }w\in W.
\end{equation*}
This is equivalent to
\begin{equation*}
\frac{\alpha b^2}{a}\in W^\perp=\F_2.
\end{equation*}
Since \(\alpha b^2/a\neq0\), this is further equivalent to
\begin{equation*}
a=\alpha b^2.
\end{equation*}
Combining the three cases proves \eqref{eq:zero-lines}.
\end{proof}

\begin{lemma}\label{lem:r-count}
Let \(q=2^m\ge4\) and fix \(\alpha\in\F_q^*\). For each
nonvertical line
\(L_{a,b}\), set
\begin{equation*}
\lambda_{a,b}=
\begin{cases}
1,&\text{if }\ell_\alpha(st)\text{ vanishes identically on }L_{a,b},\\
0,&\text{otherwise}.
\end{cases}
\end{equation*}
For \((x,y)\in\F_q^2\), define
\begin{equation*}
r(x,y)
=\#\{(a,b)\in\F_q^2\mid
b=y-ax,\ \lambda_{a,b}=1\}.
\end{equation*}
Then
\begin{equation}\label{eq:r-def}
r(x,y)=
\begin{cases}
1,&\text{if }x=0,\\[2mm]
2\left(1-\ell_\alpha(xy)\right),&\text{if }x\neq0.
\end{cases}
\end{equation}
\end{lemma}

\begin{proof}
By Lemma~\ref{lem:zero-trace-lines}, the nonzero labels among the
\(\lambda_{a,b}\) occur exactly on the horizontal line \(t=0\) and
on the \(q-1\) lines
\begin{equation*}
t=\alpha b^2s+b,\qquad b\in\F_q^*.
\end{equation*}
Therefore
\begin{equation*}
r(x,y)
=
\mathbf 1_{\{y=0\}}
+
\#\left\{b\in\F_q^*\mid
y=\alpha b^2x+b\right\}.
\end{equation*}

If \(x=0\), then
\begin{equation*}
r(0,y)
=
\mathbf 1_{\{y=0\}}
+
\#\{b\in\F_q^*\mid y=b\}
=1.
\end{equation*}

Now assume that \(x\neq0\). If \(y=0\), then
\begin{equation*}
b+\alpha b^2x=0
\end{equation*}
has exactly one solution in \(\F_q^*\), namely
\begin{equation*}
b=(\alpha x)^{-1}.
\end{equation*}
Together with the horizontal line, this gives \(r(x,0)=2\).

If $x\neq 0$ and \(y\neq0\), put \(u=\alpha xb\). Then
\begin{equation*}
y=b+\alpha b^2x
\iff
\alpha xy=u^2+u.
\end{equation*}
By Lemma~\ref{lem:W-properties}, this equation has a solution $u$ if
and only if
\begin{equation*}
\Tr_{q/2}(\alpha xy)=\ell_\alpha(xy)=0.
\end{equation*}
When it is solvable, it has exactly two solutions, and the
corresponding values of \(b=u/(\alpha x)\) are both nonzero because $x\neq 0$ and
\(y\neq0\). Hence
\begin{equation*}
r(x,y)=
\begin{cases}
2,&\ell_\alpha(xy)=0,\\
0,&\ell_\alpha(xy)=1,
\end{cases}
\end{equation*}
or equivalently
\begin{equation*}
r(x,y)=2\left(1-\ell_\alpha(xy)\right).
\end{equation*}
For \(x\neq0\) and \(y=0\), the same formula also gives
\(r(x,0)=2\), since \(\ell_\alpha(0)=0\). This proves
\eqref{eq:r-def}.
\end{proof}

\begin{lemma}\label{lem:local-weight-identity}
Let \(q=2^m\ge4\) and fix \(\alpha\in\F_q^*\). For each
vertical line
\(s=x\), set
\begin{equation*}
\mu_x=
\begin{cases}
-1,&x=0,\\
-2,&x\neq0.
\end{cases}
\end{equation*}
Then, for every \((x,y)\in\F_q^2\),
\begin{equation}\label{eq:point-identity}
\sum_{a\in\F_q}\lambda_{a,y-ax}+\mu_x
\equiv2\ell_\alpha(xy)\pmod4.
\end{equation}
\end{lemma}

\begin{proof}
If \(x=0\), then Lemma~\ref{lem:r-count} gives
\begin{equation*}
r(0,y)+\mu_0=1-1=0=2\ell_\alpha(0).
\end{equation*}
If \(x\neq0\), then
\begin{equation*}
r(x,y)+\mu_x
=2\left(1-\ell_\alpha(xy)\right)-2
\equiv2\ell_\alpha(xy)\pmod4.
\end{equation*}
Since \(r(x,y)\) is exactly the number of nonvertical lines of
label \(1\) passing through \((x,y)\), the last two displays are
exactly \eqref{eq:point-identity}.
\end{proof}

\begin{lemma}\label{lem:affine}
Let \(q=2^m\), and let
\begin{equation*}
(N_{s,t})_{s,t\in\F_q}\in\mathbb Z^{q\times q}.
\end{equation*}
Suppose that every affine line \(L\subseteq\F_q^2\) satisfies
\begin{equation*}
\sum_{(s,t)\in L}N_{s,t}\equiv0\pmod4.
\end{equation*}
Then
\begin{equation*}
\sum_{s,t\in\F_q}N_{s,t}st=0\quad\text{in }\F_q.
\end{equation*}
\end{lemma}

\begin{proof}
We first prove the assertion when \(q=2\). Every affine line of
\(\F_2^2\) contains exactly two points. Moreover, any two
distinct points \(\mathbf P,\mathbf P'\in\F_2^2\) lie on a unique affine
line. Indeed, if \(\mathbf P\neq\mathbf P'\), the nonzero vector
\(\mathbf P+\mathbf P'\) determines a unique direction, and the line
through \(\mathbf P\) with this direction is
\begin{equation*}
\{\mathbf P,\mathbf P+\mathbf P'\}=\{\mathbf P,\mathbf P'\}.
\end{equation*}
Fix a point \(\mathbf P_0\in\F_2^2\). For every
\(\mathbf P\in\F_2^2\setminus\{\mathbf P_0\}\), the line through
\(\mathbf P_0\) and \(\mathbf P\) contains exactly the two points
\(\mathbf P_0,\mathbf P\). To simplify notation, for each point
\(\mathbf P=(p_1,p_2)\in\F_2^2\), write
\begin{equation*}
N_{\mathbf P}=N_{p_1,p_2}.
\end{equation*}
Hence the hypothesis gives
\begin{equation*}
N_{\mathbf P_0}+N_{\mathbf P}\equiv0\pmod4,
\end{equation*}
or equivalently
\begin{equation*}
N_{\mathbf P}\equiv-N_{\mathbf P_0}\pmod4.
\end{equation*}
Consequently, all entries \(N_{\mathbf P}\) have the same residue modulo \(4\).
Denote this common residue by \(\lambda\). For any affine line
\(L=\{\mathbf P,\mathbf Q\}\), the hypothesis gives
\begin{equation*}
2\lambda\equiv N_{\mathbf P}+N_{\mathbf Q}\equiv0\pmod4.
\end{equation*}
Thus \(\lambda\) is even. Therefore
\begin{equation*}
\sum_{(s,t)\in\F_2^2}N_{s,t}st
=\lambda\sum_{s\in\F_2}s\sum_{t\in\F_2}t
=\lambda\cdot1\cdot1
=\lambda=0\quad\text{in }\F_2.
\end{equation*}
This proves the lemma for \(q=2\).

We now assume that \(q=2^m\ge4\). Fix
\(\alpha\in\F_q^*\). The required properties of \(W\)
are collected in Lemma~\ref{lem:W-properties}, the nonvertical lines
on which \(\ell_\alpha(st)\) vanishes identically are classified in
Lemma~\ref{lem:zero-trace-lines}, and the resulting number \(r(x,y)\)
of lines of label \(1\) through a point is determined in
Lemma~\ref{lem:r-count}. Consequently,
Lemma~\ref{lem:local-weight-identity} gives
\eqref{eq:point-identity}
for every \((x,y)\in\F_q^2\).
Multiply \eqref{eq:point-identity} by the integer \(N_{x,y}\), and sum
over all points \((x,y)\in\F_q^2\). We obtain
\begin{equation*}
2S_\alpha
\equiv
\sum_{(x,y)\in\F_q^2}N_{x,y}
\left(
\sum_{a\in\F_q}\lambda_{a,y-ax}+\mu_x
\right)
\pmod4,
\end{equation*}
where
\begin{equation*}
S_\alpha
=\sum_{x,y\in\F_q}N_{x,y}\ell_\alpha(xy)
\in\mathbb Z.
\end{equation*}
In this definition, the values of \(\ell_\alpha\) are represented by
\(0\) and \(1\).

We now rearrange the sum by lines instead of points. Every
nonvertical line \(L_{a,b}\) contains exactly the points satisfying
\(b=y-ax\), so
\begin{equation*}
\sum_{(x,y)\in\F_q^2}N_{x,y}\sum_{a\in \F_q}\lambda_{a,y-ax}
=
\sum_{(a,b)\in \F_q^2}\lambda_{a,b}\sum_{(x,y)\in L_{a,b}}N_{x,y}.
\end{equation*}
Similarly, the vertical line \(s=x\) contains exactly the points
\((x,y)\) with \(y\in\F_q\), and hence
\begin{equation*}
\sum_{x,y\in \F_q}N_{x,y}\mu_x
=
\sum_{x\in \F_q}\mu_x\sum_{y\in\F_q}N_{x,y}.
\end{equation*}
Consequently,
\begin{equation*}
2S_\alpha
\equiv
\sum_{L}\rho_L
\left(
\sum_{(x,y)\in L}N_{x,y}
\right)
\pmod4,
\end{equation*}
where \(\rho_L=\lambda_{a,b}\) if \(L=L_{a,b}\) is nonvertical and
\(\rho_L=\mu_x\) if \(L\) is the vertical line \(s=x\). By the
hypothesis of the lemma, the inner sum over every affine line \(L\)
is divisible by \(4\). Therefore
\begin{equation*}
2S_\alpha\equiv0\pmod4,
\end{equation*}
and hence \(S_\alpha\) is even.

Finally, define
\begin{equation*}
M=\sum_{x,y\in\F_q}N_{x,y}xy\in\F_q.
\end{equation*}
Since the trace is additive and takes values in \(\F_2\),
\begin{equation*}
\ell_\alpha(M)
=
\sum_{x,y\in\F_q}N_{x,y}\ell_\alpha(xy)
\equiv S_\alpha\equiv0\pmod2.
\end{equation*}
Thus \(\ell_\alpha(M)=0\) for every
\(\alpha\in\F_q^ *\). If \(M\neq0\), nondegeneracy of the
trace pairing would yield some
\(\alpha\in\F_q^*\) for which
\(\ell_\alpha(M)=1\), a contradiction. Hence \(M=0\), that is,
\begin{equation*}
\sum_{s,t\in\F_q}N_{s,t}st=0\quad\text{in }\F_q.
\end{equation*}
This completes the proof of the lemma.
\end{proof}

\begin{theorem}\label{thm:code}
Let $q=2^m$ with $m$ a positive integer, and let $\C\subseteq\F_q^n$ be a linear code over $\F_q$.
Assume that
\begin{equation*}
\one=(1,\ldots,1)\in \C
\end{equation*}
and  $\C$ is 4-divisible. 
Then
\begin{equation*}
\C\subseteq \C^{\perp_E}.
\end{equation*}
\end{theorem}

\begin{proof}
Fix $\mathbf{c}=(c_1,\ldots,c_n)\in \C$. For $s\in\F_q$, define
\begin{equation*}
n_s(\mathbf{c})=\#\{i\mid c_i=s\}.
\end{equation*}
Since $\C$ is linear, $\mathbf{c}-s\one\in \C$. The Hamming weight of
$\mathbf{c}-s\one$ is the number of positions in which $c_i-s\neq0$, or
equivalently, the number of positions in which $c_i\neq s$. Hence
\begin{equation*}
n_s(\mathbf{c})=n-\wt(\mathbf{c}-s\one).
\end{equation*}
Because $\one\in \C$ and $\one$ has weight $n$, the hypothesis gives
\begin{equation*}
n=\wt(\one)\equiv0\pmod 4.
\end{equation*}
The codeword $\mathbf{c}-s\one$ also has weight divisible by $4$. Therefore
\begin{equation}\label{eq:symbol-count}
n_s(\mathbf{c})\equiv0\pmod 4
\qquad\text{for every }s\in\F_q.
\end{equation}

Now fix  arbitrary $\mathbf{c}=(c_1,c_2,\ldots,c_n),\mathbf{d}=(d_1,d_2,\ldots,d_n)\in \C$, and define the joint symbol-count matrix \(
N = \big( N_{s,t} \big)_{s,t \in \F_q}\), where
\begin{equation*}
N_{s,t}=\#\{i\mid c_i=s,\ d_i=t\},
\qquad s,t\in\F_q.
\end{equation*}
For $a,u\in\F_q$,
\begin{equation*}
\sum_{\substack{s,t\in\F_q\\s+at=u}}N_{s,t}
=
\#\{i\mid c_i+ad_i=u\}.
\end{equation*}
Indeed, both sides count the coordinates $i$ for which
$c_i=s$, $d_i=t$, and $s+at=u$. Since $\C$ is linear,
\begin{equation*}
\bc+a\bd-u\one\in \C,
\end{equation*}
and therefore its Hamming weight is divisible by $4$. The number of
coordinates equal to $u$ in $\mathbf{c}+a\mathbf{d}$ is $n$ minus the weight of
$\mathbf{c}+a\mathbf{d}-u\one$. Since $n\equiv0\pmod 4$, we obtain
\begin{equation}\label{eq:nonhorizontal-sums}
\sum_{\substack{s,t\in\F_q\\s+at=u}}N_{s,t}\equiv0\pmod 4.
\end{equation}

We explicitly check all affine lines in the $(s,t)$-plane. If
$a\in\F_q$, the equation $s+at=u$ is a vertical line when
$a=0$. When $a\neq0$, it is the nonvertical, nonhorizontal line
\begin{equation*}
t=a^{-1}s+a^{-1}u.
\end{equation*}
Thus \eqref{eq:nonhorizontal-sums} covers every vertical line and
every nonvertical line except the horizontal lines. A horizontal line
has the form $t=u$, and its sum is handled by
\eqref{eq:symbol-count} applied to the codeword $\mathbf{d}$.
\begin{equation}\label{eq:horizontal-sums}
\sum_{s\in\F_q}N_{s,u}
=
n_u(\mathbf{d})
\equiv0\pmod 4.
\end{equation}
Equations \eqref{eq:nonhorizontal-sums} and
\eqref{eq:horizontal-sums} show that the sum of the entries of $N$
on every affine line $L\subseteq\F_q^2$ is divisible by $4$, i.e.
\begin{equation*}
\sum_{(s,t)\in L}N_{s,t}\equiv0\pmod4.
\end{equation*}
By Lemma~\ref{lem:affine},
\begin{equation*}
\sum_{s,t\in\F_q}N_{s,t}st=0\quad\text{in }\F_q.
\end{equation*}

Finally, grouping the coordinates according to the values of
$(c_i,d_i)$ gives
\begin{equation*}
\begin{aligned}
\langle \mathbf{c},\mathbf{d}\rangle_E=\sum_{i=1}^{n}c_i d_i=\sum_{s,t\in\F_q}N_{s,t}st=0.
\end{aligned}
\end{equation*}
Since $\mathbf{c},\mathbf{d}\in \C$ were arbitrary, it follows that
$\C\subseteq \C^{\perp_E}$. This proves the theorem.
\end{proof}

\begin{remark}\label{rem:allone}
The assumption \(\one\in\C\) in Theorem~\ref{thm:code} is redundant
for \(q=2\), but essential for \(q>2\).

(i) For the case \(q=2\),
it is known from \cite{HP} that every binary \(4\)-divisible code is Euclidean self-orthogonal.
Indeed, let \(\mathbf x,\mathbf y\in\C\). Since \(\C\) is linear,
\(\mathbf x+\mathbf y\in\C\). By the \(4\)-divisibility of \(\C\),
the three weights
\[
\wt(\mathbf x),\qquad
\wt(\mathbf y),\qquad
\wt(\mathbf x+\mathbf y)
\]
are all multiples of \(4\).
For binary vectors,
\[
\langle\mathbf x,\mathbf y\rangle_E
\equiv
\frac{\wt(\mathbf x)+\wt(\mathbf y)
-\wt(\mathbf x+\mathbf y)}{2}
\pmod 2 .
\]
The right-hand side is an even integer, hence
\(\langle\mathbf x,\mathbf y\rangle_E=0\) in \(\F_2\).
Therefore \(\C\subseteq\C^{\perp_E}\),
without assuming \(\one\in\C\).

(ii) For the case \(q=2^m\), \(m\ge2\), we
choose \(\alpha\in\F_q\setminus\{0,1\}\) and set
\[
\C=\bigl\langle(1,1,1,\alpha)\bigr\rangle\subseteq\F_q^4 .
\]
Every nonzero codeword of \(\C\) has weight \(4\),
so \(\C\) is \(4\)-divisible. However, for the codeword
\(\mathbf c=(1,1,1,\alpha)\in\C\),
\[
\langle\mathbf c,\mathbf c\rangle_E
=1+1+1+\alpha^2
=1+\alpha^2\neq0,
\]
because \(\alpha\neq1\).
Hence \(\C\not\subseteq\C^{\perp_E}\).
Moreover \(\one=(1,1,1,1)\notin\C\).
If \(a(1,1,1,\alpha)=(1,1,1,1)\) for some \(a\in\F_q\),
then \(a=1\) and \(a\alpha=1\), forcing \(\alpha=1\), a contradiction.
Thus the hypothesis \(\one\in\C\) cannot be removed when \(q>2\).
\end{remark}

\begin{remark}
\label{rem:grm}
Let \(q=2^s\) and \(0\le r\le m(q-1)-1\). For
\(f\in\F_q[x_1,\dots,x_m]\), put
\(\mathbf{ev}(f)=(f(\alpha))_{\alpha\in\F_q^m}\).
The generalized Reed--Muller code
\[
\mathcal R_q(r,m)
=\bigl\{\mathbf{ev}(f)\mid \deg f\le r\bigr\}
\subseteq\F_q^{q^m}
\]
is a classical code containing \(\one=\mathbf{ev}(1)\).
Its Euclidean self-orthogonality is \emph{already known} \cite{DGM70}. The point
here is that this classical code satisfies the hypotheses of
Theorem~\ref{thm:code}, so the theorem applies to a well-known family and its
assumptions are not vacuous. Indeed, by Ax's theorem \cite{Ax64},
\(\mathcal R_q(r,m)\) is \(q^{\lceil m/r\rceil-1}\)-divisible, hence
\(4\)-divisible whenever
\[
q\ge4,\ 1\le r\le m-1,
\qquad\text{or}\qquad
q=2,\ r<m/2 .
\]
In these cases, Theorem~\ref{thm:code} recovers the known self-orthogonality
\[
\mathcal R_q(r,m)\subseteq\mathcal R_q(r,m)^{\perp_E}.
\]
\end{remark}

The following is a monomial-equivalence version of Theorem \ref{thm:code}.

\begin{corollary}\label{cor:euclidean-monomial}
Let \(q=2^m\) and let \(\C\subseteq\F_q^{n}\) be a \(4\)-divisible
linear code. If \(\C\) contains a codeword of weight \(n\), then \(\C\)
is monomially equivalent to a Euclidean self-orthogonal code.
\end{corollary}

\begin{proof}
Let \(\mathbf{u}=(u_1,u_2,\ldots,u_n)\in\C\) with \(\wt(\mathbf{u})=n\). Then
\(u_{i}\ne0\) for all \(i\). Define the invertible diagonal matrix
\[
D=\operatorname{diag}(u_{1}^{-1},\dots,u_{n}^{-1}),
\]
and put \(\C'=\C D\). By definition, \(\C'\) is monomially equivalent
to \(\C\), and
\[
\mathbf{u}D
=\bigl(u_{1}u_{1}^{-1},\dots,u_{n}u_{n}^{-1}\bigr)
=(1,\dots,1)=\one\in\C'.
\]
Since monomial transformations preserve Hamming weights, \(\C'\) is
\(4\)-divisible. By Theorem \ref{thm:code}, \(\C'\) is Euclidean
self-orthogonal. Hence \(\C\) is monomially equivalent to the Euclidean
self-orthogonal code \(\C'\).
\end{proof}

\section{Hermitian Self-Orthogonal codes from \(p\)-Divisible Codes}
In the following, we mainly investigate conditions under which a linear code over \(\F_{q^2}\) is Hermitian self-orthogonal.
Here $q=p^m$ for a prime $p$ and a positive integer $m$. In addition, we generalize the obtained criterion to the Galois self-orthogonal case.

\subsection{The Hermitian criterion}
In this subsection, we establish the  Hermitian criterion as follows, which shows that every $p$-divisible linear code over $\gf_{q^2}$ containing the all-1 codeword is Hermitian self-orthogonal.
\begin{theorem}\label{thm:main}
Let \(q=p^m\) and let \(\C\subseteq\F_{q^2}^n\) be a
linear code over \(\F_{q^2}\) with \(\one\in\C\).
If \(\C\) is \(p\)-divisible, i.e.\
\(p\mid\wt(\mathbf{c})\) for every \(\mathbf{c}\in\C\),
then \(\C\) is Hermitian self-orthogonal, i.e.\
\(\C\subseteq\C^{\perp_H}\).
\end{theorem}

\begin{proof}
For \(\mathbf{c}=(c_1,c_2,\dots,c_n)\in\C\) and \(s\in\F_{q^2}\), put
\[
N_{s}(\mathbf{c})=\#\{1\leq i \leq n\mid c_{i}=s\}.
\]
Since \(\one\in\C\), we have
\(\mathbf{c}-s\one\in\C\). We have
\[
N_{s}(\mathbf{c})
=n-\wt(\mathbf{c}-s\one)
\equiv 0\pmod p,
\]
because both \(n=\wt(\one)\) and
\(\wt(\mathbf{c}-s\one)\) are divisible by \(p\).
Therefore, by Lemma~\ref{lem:hermitian}(iv),
\[
\langle \mathbf{c},\mathbf{c}\rangle_H
=\sum_{i=1}^{n}c_{i}^{q+1}
=\sum_{s\in\F_{q^2}}N_{s}(\mathbf{c})\,s^{q+1}
=0,
\]
since \(N_{s}(\mathbf{c})\equiv0\pmod p.\)

Now let \(\mathbf{c},\mathbf{d}\in\C\). For every
\(a\in\F_{q^2}\), the codeword
\(\mathbf{c}+a\mathbf{d}\) belongs to \(\C\). Hence, by
Lemma~\ref{lem:hermitian}(i)--(iii),
\[
0=\langle \mathbf{c}+a\mathbf{d},\mathbf{c}+a\mathbf{d}\rangle_H
=\langle \mathbf{c},\mathbf{c}\rangle_H
+a\langle \mathbf{d},\mathbf{c}\rangle_H
+\overline{a}\langle \mathbf{c},\mathbf{d}\rangle_H
+a\overline{a}\langle \mathbf{d},\mathbf{d}\rangle_H.
\]
By the previous paragraph,
\(\langle \mathbf{c},\mathbf{c}\rangle_H
=\langle \mathbf{d},\mathbf{d}\rangle_H=0\).
Put \(z=\langle \mathbf{c},\mathbf{d}\rangle_H\). Then
\(\langle \mathbf{d},\mathbf{c}\rangle_H=\overline{z}\) and we have
\begin{equation}\label{eq:polar}
a\overline{z}+\overline{a}z=0
\qquad\text{for all }a\in\F_{q^2}.
\end{equation}
Taking \(a=1\) in \eqref{eq:polar} gives
\(\overline{z}+z=0\), i.e.\ \(\overline{z}=-z\).
Substituting this into \eqref{eq:polar}, we obtain
\[
(\overline{a}-a)z=0
\qquad\text{for all }a\in\F_{q^2}.
\]
Choose \(a\in\F_{q^2}\setminus\F_q\). Then
\(\overline{a}\ne a\), yielding \(z=0\).
Thus \(\langle \mathbf{c},\mathbf{d}\rangle_H=0\) for all
\(\mathbf{c},\mathbf{d}\in\C\), i.e.\
\(\C\subseteq\C^{\perp_H}\).
\end{proof}

\begin{remark}\label{rem-allone}
For \(q=2\), the condition \(\one\in\C\) in
Theorem~\ref{thm:main} is not needed. Indeed, for
\(0\ne c_i\in\F_{q^2}\) one has \(c_i^{q+1}=c_i^{3}=1\), hence
\[
\langle \mathbf{c},\mathbf{c}\rangle_H
=\#\{i\mid c_{i}\ne0\}
=\wt(\mathbf{c})\pmod 2,
\]
and \(2\)-divisibility alone gives
\(\langle \mathbf{c},\mathbf{c}\rangle_H=0\) for all
\(\mathbf{c}\in\C\). Since the second paragraph in the proof of
Theorem~\ref{thm:main} does not use \(\one\in\C\), it yields
\(\langle \mathbf{c},\mathbf{d}\rangle_H=0\) for all
\(\mathbf{c},\mathbf{d}\in\C\). Hence \(\C\) is Hermitian
self-orthogonal.

For \(q>2\), the condition \(\one\in\C\) is essential in general.
For instance, we consider the following examples. 
\begin{enumerate}
\item \(q=3\). Take \(\F_{9}=\F_{3}(\alpha)\), where
\(\alpha^{2}=-1\), and
\(\mathbf{c}=(1,1+\alpha,1)\in\F_{9}^{3}\). Then
\(\wt(\mathbf{c})=3\), so \(\C=\langle\mathbf{c}\rangle\) is
\(3\)-divisible and \(\one\notin\C\), but
\[
\langle \mathbf{c},\mathbf{c}\rangle_H
=N(1)+N(1+\alpha)+N(1)
=1+2+1=1\ne0.
\]
\item \(q=4\). Take \(\F_{16}=\F_{2}(\omega)\),
\(\omega^{4}=\omega+1\), and
\(\mathbf{c}=(1,\omega)\in\F_{16}^{2}\). Then \(\wt(\mathbf{c})=2\), so
\(\C=\langle\mathbf{c}\rangle\) is \(2\)-divisible and
\(\one\notin\C\), but
\[
\langle \mathbf{c},\mathbf{c}\rangle_H=1+\omega^{5}\ne0
\]
since \(\omega^{5}\) has order \(3\).
\end{enumerate}
\end{remark}

\begin{remark}
\label{rem:grm-hermitian}
Let \(q=p^m\) and \(1\le r\le m-1\). 
The generalized Reed--Muller code
\[
\mathcal R_{q^2}(r,m)
=\bigl\{\mathbf{ev}(f)\mid \deg f\le r\bigr\}
\subseteq\F_{q^2}^{q^{2m}}
\]
is a classical code containing \(\one=\mathbf{ev}(1)\).
Its Hermitian self-orthogonality is \emph{already known} \cite{SKR05}. The point here is that these
classical code satisfies the hypotheses of Theorem~\ref{thm:main}, so the theorem
applies to a well-known family and its assumptions are not vacuous.
Indeed, by Ax's theorem \cite{Ax64}, \(\mathcal R_{q^2}(r,m)\) is
\((q^2)^{\lceil m/r\rceil-1}\)-divisible, hence \(p\)-divisible whenever
\(r\le m-1\). In these cases, Theorem~\ref{thm:main} recovers the known Hermitian
self-orthogonality
\[
\mathcal R_{q^2}(r,m)\subseteq
\mathcal R_{q^2}(r,m)^{\perp_H}.
\]
\end{remark}

The following is a monomial-equivalence version of Theorem \ref{thm:main}.

\begin{corollary}\label{cor:monomial}
Let \(q=p^m\) and let \(\C\subseteq\F_{q^2}^n\) be a
\(p\)-divisible linear code. If \(\C\) contains a codeword of weight
\(n\), then \(\C\) is monomially equivalent to a Hermitian
self-orthogonal code.
\end{corollary}

\begin{proof}
Let \(\mathbf{u}=(u_1,u_2,\ldots,u_n)\in\C\) with \(\wt(\mathbf{u})=n\). Then
\(u_{i}\ne0\) for all \(i\). Define the invertible diagonal matrix
\[
D=\operatorname{diag}(u_{1}^{-1},\dots,u_{n}^{-1}),
\]
and put \(\C'=\C D\). By definition, \(\C'\) is monomially equivalent
to \(\C\), and
\[
\mathbf{u}D
=\bigl(u_{1}u_{1}^{-1},\dots,u_{n}u_{n}^{-1}\bigr)
=(1,\dots,1)=\one\in\C'.
\]
Since monomial transformations preserve Hamming weights, \(\C'\) is
\(p\)-divisible. Therefore, by Theorem~\ref{thm:main}, \(\C'\) is
Hermitian self-orthogonal. Hence \(\C\) is monomially equivalent to
the Hermitian self-orthogonal code \(\C'\).
\end{proof}

\subsection{The Galois criterion}
\label{subsec:galois}

Let \(q=p^m\) and \(r\ge 2\). For \(0\le \ell\le r-1\), the
\emph{\(\ell\)-Galois inner product} on \(\F_{q^r}^n\) is defined by
\[
\langle\mathbf{x},\mathbf{y}\rangle_\ell
=\sum_{i=1}^{n}x_i\,y_i^{q^\ell},
\]
and the \emph{\(\ell\)-Galois dual} of a linear code
\(\mathcal{C}\subseteq\F_{q^r}^n\) is
\[
\mathcal{C}^{\perp_\ell}
=\bigl\{\mathbf{y}\in\F_{q^r}^n:\
\langle\mathbf{y},\mathbf{c}\rangle_\ell=0
\ \text{ for all }\mathbf{c}\in\mathcal{C}\bigr\}.
\]
A linear code \(\mathcal{C}\subseteq\F_{q^r}^n\) is called
\emph{\(\ell\)-Galois self-orthogonal} if
\(\mathcal{C}\subseteq\mathcal{C}^{\perp_\ell}\), i.e.,
\[
\langle\mathbf{c},\mathbf{d}\rangle_\ell=0
\quad\text{for all }\mathbf{c},\mathbf{d}\in\mathcal{C}.
\]
For \(\ell=0\), the \(\ell\)-Galois inner product and dual reduce to the
Euclidean inner product and dual, respectively. For even \(r\) and
\(\ell=r/2\), they coincide with the Hermitian inner product and dual with
respect to the subfield \(\F_{q^{r/2}}\).

\begin{theorem}
\label{thm:galois}
Let \(\C\subseteq\F_{q^r}^n\) be a linear code with
\(\one\in\C\). If \(\C\) is \(p\)-divisible and
\(\ell\in\{1,\dots,r-1\}\), then
\[
\C\subseteq\C^{\perp_\ell},
\]
that is, $\C$ is $\ell$-Galois self-orthogonal.
\end{theorem}

\begin{proof}
For \(\mathbf c\in\C\) and \(s\in\F_{q^r}\) we have
\(\mathbf c-s\one\in\C\), hence
\[
\wt(\mathbf c-s\one)\equiv0\pmod p .
\]
Put \(N_s(\mathbf c)=\#\{i:\ c_i=s\}\). Since
\(\wt(\mathbf c-s\one)=n-N_s(\mathbf c)\) and
\(n=\wt(\one)\equiv0\pmod p\), we obtain
\begin{equation}\label{eq:Ns}
N_s(\mathbf c)\equiv0\pmod p
\qquad\text{for every }s\in\F_{q^r}.
\end{equation}
Grouping the coordinates according to their value, we get
\[
\langle\mathbf c,\mathbf c\rangle_\ell
=\sum_{i=1}^{n}c_i^{1+q^\ell}
=\sum_{s\in\F_{q^r}}N_s(\mathbf c)\,s^{1+q^\ell}=0,
\]
because each summand contains the factor \(N_s(\mathbf c)\), which is
divisible by \(p\), and the characteristic is \(p\).

Now let \(\mathbf c,\mathbf d\in\C\) and \(a\in\F_{q^r}\). Since
\(\mathbf c+a\mathbf d\in\C\), the computation above gives
\[
0=\langle\mathbf c+a\mathbf d,\mathbf c+a\mathbf d\rangle_\ell
=a\langle\mathbf d,\mathbf c\rangle_\ell
+a^{q^\ell}\langle\mathbf c,\mathbf d\rangle_\ell .
\]
Putting \(z=\langle\mathbf c,\mathbf d\rangle_\ell\), the choice \(a=1\)
yields \(\langle\mathbf d,\mathbf c\rangle_\ell=-z\), and substituting
this back gives
\[
\bigl(a^{q^\ell}-a\bigr)z=0
\qquad\text{for all }a\in\F_{q^r}.
\]
Since \(\ell\ne0\), the Frobenius map \(a\mapsto a^{q^\ell}\) is not the
identity on \(\F_{q^r}\); choosing \(a\) with \(a^{q^\ell}\ne a\) we
obtain \(z=0\), that is, \(\langle\mathbf c,\mathbf d\rangle_\ell=0\).
Hence \(\C\subseteq\C^{\perp_\ell}\).
\end{proof}

\begin{corollary}\label{cor:galois-monomial}
Let \(q=p^m\) and \(r\ge2\). Let \(\C\subseteq\F_{q^r}^n\) be a
\(p\)-divisible linear code. If \(\C\) contains a codeword of weight \(n\), then \(\C\) is
monomially equivalent to a linear code \(\C'\) with
\(\C'\subseteq\C'^{\perp_\ell}\) for every \(\ell\in\{1,\dots,r-1\}\).
\end{corollary}

\begin{proof}
Let \(\mathbf u\in\C\) with \(\wt(\mathbf u)=n\). A monomial
transformation sending \(\mathbf u\) to \(\one\) preserves the Hamming
weights and hence the \(p\)-divisibility; applying
Theorem~\ref{thm:galois} to the image gives the assertion.
\end{proof}

\begin{corollary}\label{cor:galois-simultaneous}
If \(\one\in\C\) and \(\C\) is \(p\)-divisible, then
\[
\C\subseteq\bigcap_{\ell=1}^{r-1}\C^{\perp_\ell}.
\]
\end{corollary}

\begin{remark}\label{rem:galois-special}
The restriction \(\ell\ne0\) is essential in general. In the Euclidean
case \(\ell=0\), additional hypotheses are required: if \(p\) is odd,
the \(p\)-divisibility together with \(\one\in\C\) already implies
Euclidean self-orthogonality, whereas if \(p=2\), the corresponding
sufficient condition is \(4\)-divisibility (Theorem~\ref{thm:code}
applied to \(\F_{q^r}\)). For even \(r\) and \(\ell=r/2\),
Theorem~\ref{thm:galois} reduces to the Hermitian criterion
(Theorem~\ref{thm:main}); the genuinely new cases are \(r\ge3\) and
\(\ell\notin\{0,r/2\}\).
\end{remark}

\begin{remark}\label{rem:galois-sharp}
The hypothesis \(\one\in\C\) in Theorem~\ref{thm:galois} cannot be
dropped in general. Take \(q=2\), \(r=3\), \(\ell=1\), and
$\F_8=\F_2(\alpha)$, $\alpha^3=\alpha+1$,
$\C=\bigl\langle(1,1,1,\alpha)\bigr\rangle\subseteq\F_8^4$.
For every nonzero \(a\in\F_8\), all four entries of
\(a(1,1,1,\alpha)\) are nonzero, so every nonzero codeword of
\(\C\) has weight \(4\), and hence \(\C\) is \(2\)-divisible.
Moreover \(\one\notin\C\). But for \(\mathbf c=(1,1,1,\alpha)\),
\[
\langle\mathbf c,\mathbf c\rangle_1
=1^3+1^3+1^3+\alpha^3
=1+\alpha^3=1+(\alpha+1)=\alpha\ne0,
\]
so \(\C\not\subseteq\C^{\perp_1}\).
\end{remark}

\begin{remark}\label{rem:galois-converse}
For \(\ell\ne0\), Galois self-orthogonality does not force
\(p\)-divisibility in general. For instance, over \(\F_{q^2}\) with
\(q=4\) and \(\ell=1\), let \(\zeta\in\F_4\setminus\F_2\) satisfy
\(\zeta^2+\zeta+1=0\), and put
\[
\mathbf c=(0,1,\zeta,\zeta^2),\qquad
\C=\langle\one,\mathbf c\rangle\subseteq\F_{q^2}^4.
\]
Since \(\zeta^3=1\), we have
\[
\langle\one,\one\rangle_1=4=0,
\qquad \langle\one,\mathbf c\rangle_1=1+\zeta+\zeta^2=0,
\]
\[
\langle\mathbf c,\one\rangle_1=1+\zeta+\zeta^2=0,
\qquad \langle\mathbf c,\mathbf c\rangle_1
=1+\zeta^5+\zeta^{10}=1+\zeta^2+\zeta=0.
\]
By sesquilinearity, every pair of codewords in \(\C\) has inner product
zero, so \(\C\) is \(\ell\)-Galois self-orthogonal. However,
\(\wt(\mathbf c)=3\) is not divisible by \(2\), even though
\(\one\in\C\). Thus \(p\)-divisibility is not necessary even when the
all-one condition holds. In particular, the criterion of
Theorem~\ref{thm:galois} is sufficient but not necessary.
\end{remark}

\section{New families of Euclidean and Hermitian self-orthogonal codes}
In this section, we apply Theorems~\ref{thm:code} and \ref{thm:main} to two infinite families
of cyclic codes. 

For an \([n,k]\) linear code \(\C\) over \(\F_q\), define its extended code by
\begin{equation*}
\widehat{\C}
=
\left\{
(\bc,c_{n+1}):\ \bc=(c_1,\ldots,c_n)\in\C,
\quad
c_{n+1}=-\sum_{i=1}^{n}c_i
\right\}.
\end{equation*}
When the \((n+1)\)-dimensional vector \(\mathbf{1}\notin\widehat{\mathcal{C}}\), define the augmented code of $\widehat{\mathcal{C}}$ by
$
\widetilde{\C}=\widehat{\C}+\F_q\one.
$
We call \(\widetilde{\C}\) the augmented code of the extended code \(\widehat{\C}\).

\subsection{The first family of self-orthogonal codes}
\label{subsec:cyclic-first-class}

Let $q=4$. Since \(\F_4\) has characteristic two, the additional coordinate in
\(\widehat{\C}\) is also equal to \(\sum_{i=1}^{n}c_i\).

Let \(r\ge 3\) be odd and put \(n=5(4^r-1)\).
Let \(\alpha_1\) and \(\alpha_2\) be primitive elements of \(\F_{16}\) and
\(\F_{4^r}\), respectively.  Define
\begin{equation*}
\C_r
=\left\{
\left(
\Tr_{16/4}(a\alpha_1^j)
+
\Tr_{4^r/4}(b\alpha_2^j)
\right)_{j=0}^{n-1}
\mid\ a\in\F_{16},\ b\in\F_{4^r}
\right\}.
\end{equation*}
By \cite[Theorem~4.4]{liyuefu}, \(\C_r\) is a cyclic code with parameters \(\C_r=[n,r+2]\).
Fix the ordering \(0,1,\omega,1+\omega\) of the elements of \(\F_4\), where
\(\omega^2=\omega+1\).  For \(\bc\in\C_r\), let
\(N_a(\bc)=\#\{j\mid c_j=a\}\) for \(a\in\F_4\), and write \(\operatorname{comp}(\bc)=\bigl(N_0(\bc),N_1(\bc),N_\omega(\bc),N_{1+\omega}(\bc)\bigr)\).
When $\bc$ runs through $\C$, the distribution of $\operatorname{comp}(\bc)$ is called the \emph{complete weight distribution} of $\C$.

Denote by
\begin{align*}
A&=\frac{4^{r+1}-3-4^{r-1}}{3},\\
B&=\frac{4^{r+1}-4-4^{r-1}+(-2)^{r+2}}{3},\\
C&=\frac{4^{r+1}-4-4^{r-1}+(-2)^{r+1}}{3},\\
E&=\frac{4^{r+1}-4-4^{r-1}+(-2)^r}{3}.
\end{align*}
The complete weight distribution of \(\C_r\) has been given in the following lemma.

\begin{lemma}[{\cite[Theorem~4.4]{liyuefu}}]
\label{lem:li-yue-fu}
Let \(r\ge 3\) be odd and let \(\C_r\) be the cyclic code defined above.
Then \(\C_r\) has parameters \([n,r+2]\), where
\(n=5(4^r-1)\), and its possible composition vectors and multiplicities
are summarized in
Table~\ref{tab:lyf-compositions}.
\end{lemma}

\begin{table}[!t]
\caption{Composition vectors and multiplicities of the cyclic code \(\C_r\) from \cite{liyuefu}.}
\label{tab:lyf-compositions}
\centering
\renewcommand{\arraystretch}{1.15}
\begin{tabular}{@{}c l c@{}}
\hline
Condition & \(\mathbf{comp}(\bc)\) & Multiplicity\\
\hline
all odd \(r\ge3\) & \((n,0,0,0)\) & \(1\)\\
& \(\left(4^r-1,\frac{4(4^r-1)}3,\frac{4(4^r-1)}3,\frac{4(4^r-1)}3\right)\) & \(15\)\\
& \(\left(5(4^{r-1}-1),5\cdot4^{r-1},5\cdot4^{r-1},5\cdot4^{r-1}\right)\) & \(4^r-1\)\\
\hline
\(3\nmid(r-2)\) & \((A,B,C,C), (A,C,B,C), (A,C,C,B)\) & \(n\) each\\
\hline
\(3\mid(r-2)\) & \(\bigl(A-(-2)^{r+1},C,C,C\bigr)\) & \(n\)\\
& \(\bigl(A-(-2)^r,E,E,E\bigr)\) & \(2n\)\\
\hline
\end{tabular}
\end{table}

The following theorem shows that the augmented code of the extended code of \(\mathcal{C}_r\) is both Euclidean and Hermitian self-orthogonal.

\begin{theorem}
\label{thm:cyclic-self-orthogonal}
Let \(r\ge 3\) be odd, let \(n=5(4^r-1)\), and let \(\C_r\) be as above.
Define \(\widehat{\C}_r=\{(\bc,0)\mid\ \bc\in\C_r\}\) and
\(\widetilde{\C}_r=\widehat{\C}_r+\F_4\one\) with $\one$ the all-1 vector of length $n+1$.
Then \(\widetilde{\C}_r\) is an \([n+1,r+3,d]_4\) 4-divisible code over $\gf_4$ with weight enumerator
\begin{equation*}
\begin{aligned}
A_{\widetilde{\C}_r}(x)
&=1+3x^{n+1}+15x^{3D}+45x^{n+1-D}
+ (4^r-1)x^{3R}+3(4^r-1)x^{3R-4}\\
&\quad+
\begin{cases}
3nx^{n-A}+3nx^{n+1-B}+6nx^{n+1-C},
&3\nmid(r-2),\\
2nx^{n-A+(-2)^r}+3nx^{n+1-C}\\ 
\quad+nx^{n-A+(-2)^{r+1}}+6nx^{n+1-E},
&3\mid(r-2),
\end{cases}
\end{aligned}
\end{equation*} where \(d=\frac{11(4^r-1)}3+1\), \(D=\frac{4(4^r-1)}3\) and \(R=5\cdot4^{r-1}\). 
Moreover, \(\widetilde{\C}_r\) is both Euclidean and Hermitian self-orthogonal. 
\end{theorem}

\begin{proof}
It is clear that \(\widehat{\C}_r=\{(\bc,0)\mid \bc\in\C_r\}\) is the extended code of $\widehat{\C}$.
Therefore, \(\widehat{\C}_r\) has parameters
\([n+1,r+2]\).  Since \(\one \notin\widehat{\C}_r\), the
augmented code \(\widetilde{\C}_r\) has dimension \(r+3\), and its
length is \(n+1=5(4^r-1)+1=5\cdot4^r-4\).
Moreover, \(\one\in\widetilde{\C}_r\) by definition.
For $\bc\in\C_r$ and $a\in\F_4$, we have
\(\wt(\bc+a\one_n,a)=n+1-N_a(\bc)\) for \(a\ne0\).
Note that every codeword in \(\widetilde{\C}_r\) has the form $(\bc+a\one_n,a)$. 
Then the weight enumerator of \(\widetilde{\C}_r\) directly follows from
Table~\ref{tab:lyf-compositions}.
By the weight enumerator of \(\widetilde{\C}_r\), \(\widetilde{\C}_r\) is 4-divisible. 
By Theorems~\ref{thm:code} and \ref{thm:main}, \(\widetilde{\C}_r\)  is both Euclidean and Hermitian self-orthogonal.
\end{proof}

The Euclidean dual of \(\widetilde{\C}_r\) is studied as follows, which shows that the dual is distance-optimal. 
\begin{theorem}\label{cor:first-dual-distance}
Let \(r\ge3\) be odd and \(n=5(4^r-1)\), and let
\(\widetilde{\C}_r\) be the code in
Theorem~\ref{thm:cyclic-self-orthogonal}.  Then
\(
\widetilde{\C}_r^{\perp_E}
\) has parameters $\left[n+1,\,n-r-2,\,3\right]$. 
Moreover, \(\widetilde{\C}_r^{\perp_E}\) is distance-optimal.
\end{theorem}

\begin{proof}
Put $N=n+1$. For \(1\le i\le n\), let \(\mathbf e_i\in\F_4^n\) denote
the \(i\)-th standard basis vector, that is, the vector whose only
nonzero coordinate is the \(i\)-th one and equals \(1\). Note that
\(
\dim(\widetilde{\C}_r^{\perp_E})=N-(r+3)=n-r-2 .
\) We first determine the structure of the dual code.
By definition, $\widehat{\C}_r=\{(\mathbf c,0):\mathbf c\in\C_r\}$ and
$\widetilde{\C}_r=\widehat{\C}_r+\F_4\one$, hence
\[
\widetilde{\C}_r
=\bigl\{(\mathbf c+a\one,\,a):\ \mathbf c\in\C_r,\ a\in\F_4\bigr\}.
\]
A vector $(\mathbf y,\beta)\in\F_4^{N}$ is orthogonal to
$\widetilde{\C}_r$ if and only if
\[
\sum_{i=1}^{n}y_ic_i
+a\Bigl(\sum_{i=1}^{n}y_i+\beta\Bigr)=0
\qquad\text{for all }\mathbf c\in\C_r,\ a\in\F_4 .
\]
Taking $a=0$ yields $\mathbf y\in\C_r^{\perp_E}$, and then $a=1$ yields
$\beta=\sum_{i=1}^{n}y_i$. Therefore
\begin{equation}\label{eq:dual-structure}
\widetilde{\C}_r^{\perp_E}
=\Bigl\{\bigl(\mathbf y,\textstyle\sum_{i=1}^{n}y_i\bigr):
\ \mathbf y\in\C_r^{\perp_E}\Bigr\},
\end{equation}
and for every $\mathbf y\in\C_r^{\perp_E}$,
\begin{equation}\label{eq:wt-dual}
\operatorname{wt}\bigl(\mathbf y,\textstyle\sum_{i=1}^n y_i\bigr)
=\operatorname{wt}(\mathbf y)
+\Bigl[\textstyle\sum_{i=1}^{n} y_i\ne0\Bigr].
\end{equation}
Here and subsequently, \([P]\) denotes the indicator of a property
\(P\), that is, \([P]=1\) if \(P\) holds and \([P]=0\) otherwise.

We next show that $\widetilde{\C}_r^{\perp_E}$ has no nonzero codeword of
weight less than $3$. By \eqref{eq:wt-dual}, a codeword of weight $1$
would require $\operatorname{wt}(\mathbf y)=1$ and
$\sum_{i=1}^{n} y_i=0$, which is impossible. A codeword of weight $2$ occurs in one
of the following two ways.

\emph{Case 1: $\operatorname{wt}(\mathbf y)=1$ and
$\sum_{i=1}^{n} y_i\ne0$.}
Write $\mathbf y=a\mathbf e_i$ with $a\ne0$. Since
$\mathbf y\in\C_r^{\perp_E}$, we have $ac_i=0$ for all $\mathbf c=(c_1,c_2,\ldots,c_n)\in\C_r$,
i.e., the $i$-th coordinate functional vanishes on $\C_r$. But
\[
c_i=\Tr_{16/4}(a_1\alpha_1^{i-1})+\Tr_{4^r/4}(b_1\alpha_2^{i-1}),
\]
and choosing $b_1=0$ together with $a_1$ satisfying
$\Tr_{16/4}(a_1\alpha_1^i)\ne0$ gives $c_i\ne0$, a contradiction.

\emph{Case 2: $\operatorname{wt}(\mathbf y)=2$ and
$\sum_{i=1}^{n} y_i=0$.}
Write $\mathbf y=a\mathbf e_i+b\mathbf e_j$ with $i\ne j$, $a,b\ne0$
and $a+b=0$. Then $c_i=c_j$ for all $\mathbf c\in\C_r$. Substituting
the trace expressions for $c_i$ and $c_j$ and using the nondegeneracy
of the two trace forms separately, we obtain
\[
\alpha_1^i=\alpha_1^j,
\qquad
\alpha_2^i=\alpha_2^j,
\]
that is, $15\mid(i-j)$ and $(4^r-1)\mid(i-j)$. Since $r$ is odd,
$4^r\equiv1\pmod3$ and $4^r\not\equiv1\pmod5$, so
$\gcd(15,4^r-1)=3$ and
\[
\operatorname{lcm}(15,\,4^r-1)
=\frac{15(4^r-1)}{3}=5(4^r-1)=n .
\]
Thus $n\mid(i-j)$, contradicting $0<|i-j|<n$. Hence
\begin{equation}\label{eq:dual-lb}
d\bigl(\widetilde{\C}_r^{\perp_E}\bigr)\ge3 .
\end{equation}

It remains to prove $d\bigl(\widetilde{\C}_r^{\perp_E}\bigr)=3$ and  distance optimality of $\widetilde{\C}_r^{\perp_E}$.  Suppose that a quaternary
code with the same length and dimension had distance at least
\(4\).  Its parity-check matrix has \(r+3\) rows, and any three of
its columns are linearly independent.  Hence its columns define a
cap in \(\mathrm{PG}(r+2,4)\).  By Lemma~\ref{lem:barlotti-cap},
such a cap has fewer than \(4^{r+1}\) points.  On the other hand,
the number of columns is
\(n+1=5(4^r-1)+1=5\cdot4^r-4>4^{r+1}\) for \(r\ge3\),
a contradiction.  Thus no code with these parameters has distance
at least \(4\).  Hence $d\bigl(\widetilde{\C}_r^{\perp_E}\bigr)=3$ by (\ref{eq:dual-lb}).  Since $\widetilde{\C}_r^{\perp_E}$ has distance
\(3\), it is distance-optimal.
\end{proof}

\begin{example}
\label{ex:cyclic-codes}
For \(r=3\), Theorem~\ref{thm:cyclic-self-orthogonal} gives the
linear code \(\widetilde{\C}_3\) with parameters $[316,6,232]$
and Hamming weight
enumerator 
\begin{equation*}
\begin{aligned}
A_{\widetilde{\C}_3}(x)
&=
1+1935x^{232}+1134x^{236}+63x^{240}\\
&\quad+945x^{248}+15x^{252}+3x^{316}.
\end{aligned}
\end{equation*}
For \(r=5\),  Theorem~\ref{thm:cyclic-self-orthogonal} gives  \(\widetilde{\C}_5\) with parameters $[5116,8,3752]$
and Hamming weight enumerator 
\begin{equation*}
\begin{aligned}
A_{\widetilde{\C}_5}(x)
&=
1+45x^{3752}+10230x^{3804}+15345x^{3816}\\
&\quad+3069x^{3836}+1023x^{3840}
+30690x^{3848}\\
&\quad+5115x^{3900}+15x^{4092}+3x^{5116}.
\end{aligned}
\end{equation*}
In both cases, all nonzero weights are divisible by \(4\). Hence both \(\widetilde{\C}_3\) and \(\widetilde{\C}_5\) are Euclidean and Hermitian self-orthogonal by Theorems \ref{thm:code} and \ref{thm:main}. 
\end{example}

\subsection{The second family of self-orthogonal codes}
\label{subsec:cyclic-second-class}

We next consider another infinite family of cyclic codes.
The construction is given in the following lemma.

\begin{lemma}
\label{lem:xiong-family}
Let \(q=2^m\), \(m\ge2\), and \(r\ge2\), and put
\(N=q^r-1\). Let \(\gamma\) be a primitive element of
\(\F_{q^r}\), and define
\begin{equation}
\label{eq:second-cyclic-code}
\C_{q,r}
=
\left\{
\left(
\Tr_{q^r/q}\left(a\gamma^{i}+b\gamma^{2i}\right)
\right)_{i=0}^{N-1}
\mid
a,b\in\F_{q^r}
\right\}.
\end{equation}
Then \(\C_{q,r}\) is a cyclic code over \(\F_q\) with parameters
\([q^r-1,2r]\).
\end{lemma}

\begin{proof}
A cyclic shift replaces \((a,b)\) by
\((a\gamma^{-1},b\gamma^{-2})\), so it preserves \(\C_{q,r}\).
Hence \(\C_{q,r}\) is cyclic.

We next prove that the trace representation is injective. Suppose
that \(a,b\in\F_{q^r}\) give the zero codeword, and put
\(P(z)=\Tr_{q^r/q}(az+bz^{2})\). Since \(\gamma\) is primitive,
\(P(z)=0\) for every \(z\in\F_{q^r}^{*}\). Expanding the two
traces gives
\(
P(z)=\sum_{j=0}^{r-1}
\left(a^{q^{j}}z^{q^{j}}+b^{q^{j}}z^{2q^{j}}\right).
\)
Because \(q\ge4\), the exponent \(2q^{j}\) is never of the form
\(q^{i}\). Also,
\(\deg P\le2q^{r-1}<q^r-1=N\). Thus \(P\) has more roots than its
degree, so it is the zero polynomial. The coefficients of \(z\) and
\(z^{2}\) in the displayed expansion are \(a\) and \(b\), respectively.
Therefore \(a=b=0\), and the representation is injective. Its image
therefore has dimension \(2r\) over \(\F_q\).

Consequently, \(\C_{q,r}\) is a cyclic code with parameters
\([q^r-1,2r]\).  We note that this trace-code construction
is the case \(t=2\) of the family of reducible cyclic codes
considered in \cite[Section II]{xiongetal}.
\end{proof}

We now apply the extending and augmentation technique to obtain self-orthogonal codes from \(\C_{q,r}\).

\begin{theorem}
\label{thm:xiong-cyclic-self-orthogonal}
Let \(q=2^m\), \(m\ge2\), and \(r\ge2\). Let \(N=q^r-1\),
and define \(\widehat{\C}_{q,r}=\{(\bc,0)\mid \bc\in\C_{q,r}\}\) and
\(\widetilde{\C}_{q,r}=\widehat{\C}_{q,r}+\F_q\one\) with $\one$ the all-vector of length $N+1$.
Then \(\widetilde{\C}_{q,r}\) is a $q$-divisible \([q^r,2r+1]\) linear code with weight enumerator
\begin{equation*}
\begin{aligned}
A_{\widetilde{\C}_{q,r}}(x)
={}&1
+\frac{q(q-1)}2(q^r-1)x^{q^{r-1}(q-2)}+q(q^r-1)(q^r-q+2)x^{q^{r-1}(q-1)}\\
&+\left(\frac{q(q-1)}2(q^r-1)+q-1\right)x^{q^r}.
\end{aligned}
\end{equation*}
Besides, $\widetilde{\C}_{q,r}$ is Euclidean self-orthogonal; for even $m$,  $\widetilde{\C}_{q,r}$ is Hermitian self-orthogonal. 
\end{theorem}

\begin{proof}
We divide the proof into the following steps.

\emph{Step 1. The dimension of $\widetilde{\C}_{q,r}$.}
Since \(\one \notin\widehat{\C}_{q,r}\), it is obvious that
\(\dim(\widetilde{\C}_{q,r})=2r+1\) by definition.

\emph{Step 2. The Hamming weight enumerator.}
Denote by $\bc_{a,b}=\left(
\Tr_{q^r/q}\left(a\gamma^{i}+b\gamma^{2i}\right)\right)_{i=0}^{N-1}\in \C_{q,r}$.
For \(f(x)=\Tr_{q^r/q}(ax+bx^2)\) and $\varepsilon\in \F_q$, we have
\[
(\bc_{a,b},0)+\varepsilon\one
=(f(x)+\varepsilon)_{x\in\F_{q^r}}.
\]
Note that every $\bc\in\widetilde{\C}_{q,r}$ can be written as the form $(\bc_{a,b},0)+\varepsilon\one$. 
Put
\(
N_f(\varepsilon)=|\{x\in\F_{q^r}:f(x)=\varepsilon\}|.
\)
Then
\[
\wt\bigl((\bc_{a,b},0)+\varepsilon\one\bigr)
=q^r-N_f(\varepsilon).
\]
For \((a,b)\ne(0,0)\), \(f\) is \(\F_2\)-linear. Define
\[
(\mathrm{Im}\,f)^{\perp}
=\bigl\{c\in\F_q\mid \Tr_{q/2}(cf(x))=0\ \text{for all }
x\in\F_{q^r}\bigr\}.
\]
Since the pairing \((c,y)\mapsto\Tr_{q/2}(cy)\) on \(\F_q\times\F_q\) is
perfect, \(|(\mathrm{Im}\,f)^{\perp}|=q/|\mathrm{Im}\,f|\).
For every \(c\in\F_q^{\ast}\), 
\[
\begin{aligned}
c\in(\mathrm{Im}\,f)^{\perp}
&\Longleftrightarrow
\Tr_{q^r/2}\bigl(cax+cbx^2\bigr)=0
\quad\text{for all }x\in\F_{q^r}\\
&\Longleftrightarrow
cb=(ca)^2
\Longleftrightarrow
c\,(ca^2+b)=0
\Longleftrightarrow
b=ca^2 .
\end{aligned}
\]
Therefore
\[
(\mathrm{Im}\,f)^{\perp}
=\{0\}\cup\{c\in\F_q^{\ast}:\ b=ca^2\}.
\]
The equation \(b=ca^2\) has at most one solution in \(c\); hence
\(|(\mathrm{Im}\,f)^{\perp}|\in\{1,2\}\), and
\[
|\mathrm{Im}\,f|
=\frac{q}{|(\mathrm{Im}\,f)^{\perp}|}
\in\Bigl\{q,\ \frac q2\Bigr\}.
\]
More precisely, \(|(\mathrm{Im}\,f)^{\perp}|=2\) exactly when
\(a\ne0\) and \(c_0:=b/a^2\in\F_q^{\ast}\); in this case
\((\mathrm{Im}\,f)^{\perp}=\{0,c_0\}\).
Since \(f\) is \(\F_2\)-linear, every value in \(\mathrm{Im}\,f\) is
attained exactly \(|\ker f|=q^r/|\mathrm{Im}\,f|\) times, and every value
outside \(\mathrm{Im}\,f\) is attained \(0\) times. Put
\(W=\{w\in\F_q:\Tr_{q/2}(w)=0\}\), so that \(|W|=q/2\) and
\((\mathrm{Im}\,f)^{\perp}=\{0,c_0\}\) gives
\(\mathrm{Im}\,f=c_0^{-1}W\). Hence
\[
N_f(\varepsilon)
=\begin{cases}
2q^{r-1}, & \text{if }b=c_0a^2,\ \Tr_{q/2}(c_0\varepsilon)=0,\\[2pt]
0, & \text{if }b=c_0a^2,\ \Tr_{q/2}(c_0\varepsilon)\ne0,\\[2pt]
q^{r-1}, & \text{if }\text{otherwise},
\end{cases}
\tag{6}
\]
where \(c_0=b/a^2\in\F_q^{\ast}\) in the first two lines, and
\(\mathrm{Im}\,f=\F_q\) in the last one.
Then the weights of $\widetilde{\C}_{q,r}$ follow from $\wt\bigl((\bc_{a,b},0)+\varepsilon\one\bigr)
=q^r-N_f(\varepsilon).$
 For $(a,b)=(0,0)$, we have
\[
N_f(0)=q^r,
\qquad
N_f(\varepsilon)=0
\quad(\varepsilon\in\F_q^*).
\]
Since
\[
\#\{(a,b,c_0):b=c_0a^2,\ c_0\in\F_q^*,
\ a\in\F_{q^r}^*\}=(q-1)(q^r-1),
\]
and
\[
q^{2r}-1-(q-1)(q^r-1)=(q^r-1)(q^r-q+2),
\]
then we deduce
\begin{align*}
\#\{\bc\in\widetilde{\C}_{q,r}:\wt(\bc)=q^{r-1}(q-2)\}
&=\frac{q(q-1)}2(q^r-1),\\
\#\{\bc\in\widetilde{\C}_{q,r}:\wt(\bc)=q^{r-1}(q-1)\}
&=q(q^r-1)(q^r-q+2),\\
\#\{\bc\in\widetilde{\C}_{q,r}:\wt(\bc)=q^r\}
&=\frac{q(q-1)}2(q^r-1)+q-1,
\end{align*}
which gives the stated Hamming weight enumerator.

\emph{Step 3. Applications of the self-orthogonality criteria.}
Note that \(\widetilde{\C}_{q,r}\) is linear over \(\F_q\), contains
\(\one\), and is \(q\)-divisible. Since \(q=2^m\) with
\(m\ge2\), Theorem~\ref{thm:code} gives
\(\widetilde{\C}_{q,r}\subseteq
\widetilde{\C}_{q,r}^{\perp_E}\).
If \(m\) is even, then \(\F_q=\F_{(2^{m/2})^2}\). The code is
linear over this square field and contains \(\one\).  Applying
Theorem~\ref{thm:main} 
gives \(\widetilde{\C}_{q,r}\subseteq\widetilde{\C}_{q,r}^{\perp_H}\)
in \(\F_q^{N+1}\), with conjugation \(x\mapsto x^{2^{m/2}}\). This proves the theorem.
\end{proof}

\begin{theorem}\label{cor:second-dual-distance}
Let \(q=2^m\), \(m\ge2\), and \(r\ge2\). Then \(\widetilde{\C}_{q,r}^{\perp_E}\) has parameters $\left[q^r,\,q^r-2r-1,\,4\right]$.
Moreover, \(\widetilde{\C}_{q,r}^{\perp_E}\) is distance-optimal.
\end{theorem}

\begin{proof}

Put \(n=q^r\) and \(k=2r+1\). Let \(B_j=A_j^\perp\), where \(B_j\) denotes the frequency of codewords of weight \(j\) in \(\widetilde{\mathcal{C}}_{q,r}^{\perp_E}\). By
Theorem~\ref{thm:xiong-cyclic-self-orthogonal}, the only
nonzero weights of \(\widetilde{\mathcal{C}}_{q,r}\) are
\[
w_1=q^{r-1}(q-2),\qquad
w_2=q^{r-1}(q-1),\qquad
w_3=n,
\]
with frequencies
\[
A_{w_1}=\frac{q(q-1)}2(q^r-1),\quad
A_{w_2}=q(q^r-1)(q^r-q+2),
\quad A_{w_3}=\frac{q(q-1)}2(q^r-1)+q-1.
\]
By the \(q\)-ary MacWilliams identity (equivalently, the Pless power
moments) \cite[pp.~260--261]{HP},
\begin{equation*}
B_i=q^{-k}\sum_{j=0}^{n}A_jK_i(j),\qquad
K_i(j)=\sum_{s=0}^{i}(-1)^s(q-1)^{i-s}
\binom{j}{s}\binom{n-j}{i-s}.
\end{equation*}
Substitution gives
\begin{align*}
B_1=B_2=B_3=0,\
B_4=\frac{n(n-1)(q-1)(n+q^2-6q+6)}{24}>0.
\end{align*}
Thus the dual distance
is exactly \(4\).

It remains to prove distance optimality.  We use the \(q\)-ary
Hamming bound and include its short proof.  Put \(N=q^r\), and
suppose that a \(q\)-ary linear code with the same length and
dimension had distance at least \(5\). Then
\[
q^{N-2r-1}\left(1+N(q-1)+\binom N2(q-1)^2\right)
\le q^N.
\]
After division by \(q^N\), this gives
\(1+N(q-1)+\binom N2(q-1)^2\le q^{2r+1}=qN^2\).
However, \(N=q^r\ge16\) and \(q\ge4\), so
\[
\frac{\binom N2(q-1)^2}{qN^2}
=\frac{(N-1)(q-1)^2}{2qN}
\ge\frac{15}{16}\cdot\frac94\cdot\frac12
=\frac{135}{128}>1.
\]
Consequently, \( qN^2 < \binom{N}{2}(q-1)^2 \), which is a contradiction. Thus no
code with these parameters has distance at least \(5\).  Since
\(\widetilde{\C}_{q,r}^{\perp_E}\) has distance \(4\), it is
distance-optimal.
\end{proof}

The localities of \(\widetilde{\C}_{q,r}\)  and \(\widetilde{\C}_{q,r}^{\perp_E}\) are studied as follows.

\begin{proposition}\label{prop:second-lrc}
Let \(q=2^m\), \(m\ge2\) and \(r\ge2\). Then $\widetilde{\C}_{q,r}$ has locality \(3\), while
\(\widetilde{\C}_{q,r}^{\perp_E}\) has locality \(q^{r-1}(q-2)-1\). In the
special case \(r=2\), the dual code \(\widetilde{\C}_{q,r}^{\perp_E}\) is an almost
distance-optimal LRC.
\end{proposition}

\begin{proof}
By Theorem~\ref{cor:second-dual-distance},
\(d(\widetilde{\C}_{q,r}^{\perp_E})=4\). Combining Theorem \ref{thm:xiong-cyclic-self-orthogonal} and Lemma \ref{lem:assmus-mattson} shows that the
minimum-weight words of \(\widetilde{\C}_{q,r}^{\perp_E}\) form a \(2\)-design. Applying
Lemma~\ref{lem:dual-design-locality} gives that $\widetilde{\C}_{q,r}$ has
locality \(4-1=3\). Similarly,
\(d(\widetilde{\C}_{q,r})=q^{r-1}(q-2)\) by
Theorem \ref{thm:xiong-cyclic-self-orthogonal}, and the minimum-weight
words of \(\widetilde{\C}_{q,r}\) form a \(2\)-design by
Theorem \ref{thm:xiong-cyclic-self-orthogonal} and Lemma \ref{lem:assmus-mattson}.
Thus the locality of \(\widetilde{\C}_{q,r}^{\perp_E}\) is
\(d(\widetilde{\C}_{q,r})-1=q^{r-1}(q-2)-1\).

It remains to prove the assertions for \(r=2\). In this case,
\(\widetilde{\C}_{q,r}^{\perp_E}\) has parameters
\([q^2,q^2-5,4]\) and locality
\(\ell=q(q-2)-1\). The
Singleton-like bound in Lemma~\ref{lem:lrc-singleton} shows \(\widetilde{\C}_{q,r}^{\perp_E}\) is an almost
distance-optimal LRC.
\end{proof}

\begin{example}
\label{ex:xiong-cyclic-codes}
For \((q,r)=(4,2)\), Theorem
\ref{thm:xiong-cyclic-self-orthogonal} gives the linear code
\(\widetilde{\C}_{4,2}\) with parameters $[16,5,8]$ and weight enumerator
\[A_{\widetilde{\C}_{4,2}}(x)
=1+90x^{8}+840x^{12}+93x^{16}.\]
For \((q,r)=(8,2)\), Theorem
\ref{thm:xiong-cyclic-self-orthogonal}  gives the linear code
\(\widetilde{\C}_{8,2}\) with parameters $[64,5,48]$ and weight enumerator
\[A_{\widetilde{\C}_{8,2}}(x)
=1+1764x^{48}+29232x^{56}+1771x^{64}.\]
In both cases all nonzero weights are divisible by \(4\). Then both \(\widetilde{\C}_{4,2}\) and \(\widetilde{\C}_{8,2}\) are Euclidean self-orthogonal by Theorem \ref{thm:code}.
Besides, \(\widetilde{\C}_{4,2}\) is also Hermitian self-orthogonal by Theorem \ref{thm:main}. 
\end{example}

\section{Quantum codes from self-orthogonal codes}
\label{sec:quantum-codes}

We briefly recall the terminology used below. A \(q\)-ary quantum
code with parameters \([[n,k,d]]_q\) is a \(q^k\)-dimensional
subspace of \((\bC^q)^{\otimes n}\), where \(d\) is its minimum
distance. The Hermitian and CSS constructions below both yield
stabilizer quantum codes \cite{Ketkar06,FengMa04}. We say that such
a code is pure if its stabilizer group contains no non-scalar error
of weight less than \(d\). Otherwise, it is called impure.

\begin{lemma}\cite[Corollary~19, Hermitian construction]{Ketkar06}
\label{lem:hermitian-construction}
Let \(\C=[n,k,d]\) be a linear code over $\gf_{q^2}$ satisfying
\(\C\subseteq\C^{\perp_H}\). Put \(d_H=d(\C^{\perp_H})\).
Then there exists a \(q\)-ary quantum code with parameters
\[
[[n,\,n-2k,\,\ge d_H]]_q.
\]
In particular, if \(d_H<d\), then its minimum distance is exactly
\(d_H=d(\C^{\perp_H}\setminus\C)\).
\end{lemma}

\begin{lemma}\cite[Theorem~27]{Ketkar06}
\label{lem:quantum-hamming-two-errors}
If a \(q\)-ary stabilizer code with  parameters \([[n,k,5]]_q\) exists, then
\[
1+n(q^2-1)+\binom n2(q^2-1)^2\le q^{n-k}.
\]
\end{lemma}

\begin{theorem}\label{thm:hermitian-quantum-second-class}
Let \(q=2^m\), where \(m\ge2\) is even. Then there exists
a \(2^{m/2}\)-ary distance-optimal quantum code
with parameters
\[
[[q^r,\,q^r-4r-2,\,4]]_{2^{m/2}}.
\]
\end{theorem}

\begin{proof}
Since \(q=(2^{m/2})^2\), applying
Lemma~\ref{lem:hermitian-construction} with base field
\(\F_{2^{m/2}}\) gives a \(2^{m/2}\)-ary quantum code
with parameters
\[
[[q^r,\,q^r-2(2r+1),\,4]]_{2^{m/2}}
=[[q^r,\,q^r-4r-2,\,4]]_{2^{m/2}}
\]
by Theorem \ref{thm:xiong-cyclic-self-orthogonal} and Theorem \ref{cor:second-dual-distance}.
This quantum code has alphabet
size \(2^{m/2}\), and
$n=q^r=2^{mr}$, $k=n-4r-2$.
Suppose that a \(2^{m/2}\)-ary stabilizer code with the same length
and dimension had distance at least \(5\). By
Lemma~\ref{lem:quantum-hamming-two-errors},
\[
1+n(2^m-1)+\binom n2(2^m-1)^2
\le(2^{m/2})^{n-k}=(2^{m/2})^{4r+2}=2^mn^2.
\]
On the other hand, \(m\ge2\) and \(r\ge2\), so \(n\ge16\) and
\(2^m-1\ge3\). Therefore
\[
\frac{n(n-1)(2^m-1)^2/2}{2^mn^2}
=\frac{(n-1)(2^m-1)^2}{2^{m+1}n}
>1.
\]
Consequently,
\[
1+n(2^m-1)+\binom n2(2^m-1)^2
\ge\frac{n(n-1)}2(2^m-1)^2>2^mn^2,
\]
which contradicts the double-error quantum Hamming bound.
Hence no stabilizer code with the same alphabet, length and dimension has
distance at least \(5\). This completes the proof.
\end{proof}

\begin{lemma}\cite[Lemma~20, CSS construction]{Ketkar06}]
\label{lem:css}
Let \(\C=[n,k,d]\) be a linear code over $\gf_q$ with
\(\C\subseteq\C^{\perp_E}\), and put
\(d_Q=d(\C^{\perp_E}\setminus\C)\). Then there exists a \(q\)-ary
CSS stabilizer code with parameters
\[
[[n,\,n-2k,\,d_Q]]_q.
\]
Moreover, if \(d_Q=d(\C^{\perp_E})\), then it is pure. In
particular, if \(d(\C^{\perp_E})<d\), then
\(d_Q=d(\C^{\perp_E})\) and the code is pure.
\end{lemma}

\begin{theorem}\label{thm:css-first-class}
Let \(r\ge3\) be odd. Then there
exists a quaternary CSS quantum code with parameters
\[
[[5\cdot4^r-4,\,5\cdot4^r-2r-10,\,3]]_4.
\]
\end{theorem}

\begin{proof}
Applying
Lemma~\ref{lem:css} gives a quaternary CSS quantum code with
parameters
\([[5\cdot4^r-4,5\cdot4^r-2r-10,3]]_4\)  by Theorem \ref{thm:cyclic-self-orthogonal} and Theorem \ref{cor:first-dual-distance}.
\end{proof}

\begin{theorem}\label{thm:css-second-class}
Let \(q=2^m\), \(m\ge2\), and \(r\ge2\). Then there exists
a \(q\)-ary CSS quantum code with parameters
\[
[[q^r,\,q^r-4r-2,\,4]]_q.
\]
\end{theorem}

\begin{proof}
Applying
Lemma~\ref{lem:css} gives the stated \(q\)-ary CSS quantum
code by Theorem \ref{thm:xiong-cyclic-self-orthogonal} and Theorem \ref{cor:second-dual-distance}.
\end{proof}

\section{Concluding remarks}
In this paper, we addressed Question~1 and Question~2 by establishing criteria for self-orthogonal codes derived from divisible codes. 
The Euclidean criterion (Theorem~\ref{thm:code}) generalizes Huffman and Pless's result from $q=2$ to $q=2^m$ through an additional condition. The Hermitian criterion (Theorem~\ref{thm:main}) generalizes Huffman and Pless's result from $q=2$ to $q=p^m$ through an additional condition. These criteria hold considerable potential for constructing new families of Euclidean and Hermitian self-orthogonal codes, since divisible codes containing the all-1 vector are very easy to construct. In addition to the divisible codes presented in this paper, further such divisible codes can be found in~\cite{liheng}. 
In addition, we generalized the Hermitian  criterion to the Galois self-orthogonal case.

\section*{Declarations}
\textbf{Conflict of interest} The authors declare that they have no conflicts of interest relevant to the content of this article.



\begin{thebibliography}{99}

\bibitem{AM69}
E. F. Assmus Jr., H. F. Mattson Jr., New \(5\)-designs, \emph{J. Combin. Theory} \textbf{6} (1969) 122--151.

\bibitem{Ax64}
J. Ax, Zeroes of polynomials over finite fields, \emph{Amer. J. Math.} \textbf{86} (1964) 255--261.

\bibitem{Barlotti}
A. Barlotti, Bounds for \(k\)-caps in \(\mathrm{PG}(r,q)\) useful in the theory of error correcting codes, Institute of Statistics Mimeo Series No.~484.2, University of North Carolina, 1966.

\bibitem{CRSS98}
A. R. Calderbank, E. M. Rains, P. W. Shor, N. J. A. Sloane, Quantum error correction via codes over GF(4), \emph{IEEE Trans. Inform. Theory} \textbf{44} (1998) 1369--1387.

\bibitem{DGM70}
P. Delsarte, J.-M. Goethals, F. J. MacWilliams, On generalized Reed--Muller codes and their relatives, \emph{Inform. and Control} \textbf{16} (1970) 403--442.

\bibitem{Ding}
C. Ding, C. Tang, \emph{Designs from Linear Codes}, 2nd ed., World Scientific, Singapore, 2022.

\bibitem{FengMa04}
K. Feng, Z. Ma, A finite Gilbert--Varshamov bound for pure stabilizer quantum codes, \emph{IEEE Trans. Inform. Theory} \textbf{50} (2004) 3323--3325.

\bibitem{Gopalan12}
P. Gopalan, C. Huang, H. Simitci, S. Yekhanin, On the locality of codeword symbols, \emph{IEEE Trans. Inform. Theory} \textbf{58} (2012) 6925--6934.

\bibitem{HP}
W. C. Huffman, V. Pless, \emph{Fundamentals of Error-Correcting Codes}, Cambridge University Press, Cambridge, 2003.

\bibitem{Ketkar06}
A. Ketkar, A. Klappenecker, S. Kumar, P. K. Sarvepalli, Nonbinary stabilizer codes over finite fields, \emph{IEEE Trans. Inform. Theory} \textbf{52} (2006) 4892--4914.

\bibitem{KK}
M. Kiermaier, S. Kurz, On the lengths of divisible codes, \emph{IEEE Trans. Inform. Theory} \textbf{66} (2020) 4051--4060.

\bibitem{liyuefu}
C. Li, Q. Yue, F.-W. Fu, Complete weight enumerators of some cyclic codes, \emph{Des. Codes Cryptogr.} \textbf{80} (2016) 295--315.

\bibitem{liheng}
X. Li, Z. Heng, Self-orthogonal codes from \(p\)-divisible codes, \emph{IEEE Trans. Inform. Theory} \textbf{70} (2024) 8562--8586.

\bibitem{Massey}
J. L. Massey, Orthogonal, antiorthogonal and self-orthogonal matrices and their codes, in: \emph{Communications and Coding}, Research Studies Press, Somerset, U.K., 1998.

\bibitem{SKR05}
P. K. Sarvepalli, A. Klappenecker, Nonbinary quantum Reed--Muller codes, in: \emph{Proc. IEEE Int. Symp. Inform. Theory (ISIT)}, Adelaide, Australia, 2005, pp. 1023--1027.

\bibitem{Tan23}
P. Tan, C. Fan, C. Ding, C. Tang, Z. Zhou, The minimum locality of linear codes, \emph{Des. Codes Cryptogr.} \textbf{91} (2023) 83--114.

\bibitem{Wan}
Z. Wan, A characteristic property of self-orthogonal codes and its application to lattices, \emph{Bull. Belg. Math. Soc. Simon Stevin} \textbf{5} (1998) 477--482.

\bibitem{WangHeng24}
X. Wang, Z. Heng, Several families of self-orthogonal codes and their applications in optimal quantum codes and LCD codes, \emph{IEEE Trans. Inform. Theory} \textbf{70} (2024) 4769--4791.

\bibitem{Ward1}
H. N. Ward, Divisible codes, \emph{Arch. Math.} \textbf{36} (1981) 485--494.

\bibitem{Ward2}
H. N. Ward, Divisible codes---a survey, \emph{Serdica Math. J.} \textbf{27} (2001) 263--278.

\bibitem{xiongetal}
M. Xiong, S. Li, G. Ge, The weight hierarchy of some reducible cyclic codes, \emph{IEEE Trans. Inform. Theory} \textbf{62} (2016) 4071--4080.

\end{thebibliography}
\end{document}